\documentclass[a4paper,11pt]{article}
\pdfoutput=1
\usepackage{jcappub}

\usepackage[applemac]{inputenc}
\usepackage{graphicx}
\usepackage{amsmath}
\usepackage{amsfonts}
\usepackage{amssymb}
\usepackage{makeidx}
\usepackage{wrapfig}
\usepackage{empheq}
\usepackage{geometry}
\usepackage{comment}
\usepackage{soul}
\usepackage{ulem}

\def\beq{\begin{equation}}
\def\eeq{\end{equation}}
\def\bea{\begin{eqnarray}}
\def\eea{\end{eqnarray}}
\def\bean{\begin{eqnarray*}}
\def\eean{\end{eqnarray*}}

\def \dd {\partial}

\def \La {\Lambda}

\def \De {\Delta}

\def \Om {\Omega}

    \def\be{\begin{equation}}
    \def\ee{\end{equation}}
    \def\ba{\begin{eqnarray}}
    \def\ea{\end{eqnarray}}

    \newcommand{\bn}{{\bf n}}
    \newcommand{\NN}{{\mathcal{N}}}
    \newcommand{\HH}{{\mathcal{H}}}
    \newcommand{\ndv}{\bn \cdot {\bf v}}
    \newcommand{\bv}{ {\bf v}}
        \newcommand{\bk}{ {\bf k}}

\newcommand{\class}{{\sc class}}
\newcommand{\classgal}{{\sc class}gal}

\newcommand{\lya}{Lyman-$\alpha$}
\newcommand{\elya}{\mathrm{Ly}\alpha}

\usepackage{color}

\begin{document}

\title{Relativistic effects in Lyman-$\alpha$ forest}

\author[a]{Vid Ir\v{s}i\v{c},}
\author[b,c]{Enea Di Dio,}
\author[b,c]{Matteo Viel}

\affiliation[a]{The Abdus Salam International Centre for Theoretical
  Physics, Strada Costiera 11, I-34151 Trieste, Italy}
\affiliation[b]{INAF - Osservatorio Astronomico di Trieste, Via
  G. B. Tiepolo 11, I-34143 Trieste, Italy}
\affiliation[c]{INFN-National Institute for Nuclear Physics,
via Valerio 2, I-34127 Trieste, Italy}

\emailAdd{virsic@ictp.it}
\emailAdd{enea.didio@oats.inaf.it}
\emailAdd{viel@oats.inaf.it}

\date{\today}

\abstract{
We present the calculation of the Lyman-alpha (\lya) transmitted flux fluctuations with
full relativistic corrections to the first order. Even though several
studies exist on relativistic effects in galaxy clustering, this
is the first study to extend the formalism to a different tracer of
underlying matter at unique redshift range ($z=2-5$). Furthermore, we show a
comprehensive application of our calculations to the
Quasar-\lya\ cross-correlation function. Our results indicate that the
signal of relativistic effects are sizeable at Baryonic Acoustic
Oscillation (BAO) scale mainly due to the large
differences in density bias factors of our tracers. We construct an
observable, the anti-symmetric part of the cross-correlation function,
that is dominated by the relativistic signal and offers a new way to
measure the relativistic terms at relatively small scales. The
analysis shows that relativistic effects are important when
considering cross-correlations between tracers with very different
biases, and should be included in the data analysis of the current and
future surveys. Moreover, the idea presented in this paper is highly
complementary to other techniques and observables trying to isolate
the effect of the relativistic corrections and thus test the validity
of the theory of gravity beyond the Newtonian regime.
}

\maketitle

\section{Introduction}

The last decade has witnessed an enormous progress in the cosmological
investigation of the intergalactic medium (IGM) as probed by the
\lya\ forest (see \cite{rauch98,meiksin09} for reviews). In
particular, the \lya\ forest is now a viable cosmological observable
that can help in putting constraints on cosmological parameters and/or
deviations from a standard scenario based on Cold Dark Matter (CDM) and a
cosmological constant. More recently, the very
large number of quasar spectra of the SDSS-III/BOSS (Sloan Digital
Sky Survey-III/Baryon Oscillation Spectroscopic Survey) collaboration
using a state-of-the-art analysis, has allowed to discover
Baryonic Acoustic Oscillations (BAOs) in the transmitted \lya\ flux at
$z\sim 2.3$ \cite{busca13,slosar13}: this spectacular
confirmation of the cosmological nature of the \lya\ forest was mainly
made possible by exploiting the flux correlations in three dimensions
out to large scales.  Moreover, the one-dimensional analysis of the transmitted
flux power has allowed to place the most stringent bounds on total neutrino mass and cold dark matter coldness
by using spectra at high and low resolution \cite{palanque13,palanque15,viel13}.

The first steps have also been taken in measuring the
cross-correlations between various different observables in the same
volume of the survey. Such examples include the cross-correlations
between the \lya\ forest and Damped Lyman-$\alpha$ systems
 \cite{font12DLA} and the \lya\ forest and Quasars
(QSOs)~\cite{font13,font14}. The first measurements measured the correlations
up to the BAO scale and confirmed the location of the peak, however the
error bars remain large. Nevertheless, the analysis was done on barely
half of the available data by the SDSS-III/BOSS survey and future
generation of large-scale structure surveys (e.g.~SDSS-IV/eBOSS~\cite{eBOSS},
DESI~\cite{DESI}, etc.) are going to improve on the precision of the measurements
to a percent level thus forcing the theoretical models to match unprecedented demanding constraints
in terms of accuracy.

However, as the observations move out to larger scales the validity of
a simple Newtonian description starts to come into question. Recent
years have seen a lot of effort put into understanding and possibly
measuring the signal of relativistic effects and/or weak gravitational
lensing in the 2-point statistics of the galaxy clustering, see
e.g.~\cite{McDonald:2009ud,Yoo:2012se,Yoo:2013zga,croft13,DiDio:2013sea,Bonvin:2013ogt,Alonso:2015uua,Alonso:2015sfa,Fonseca:2015laa}. Moreover,
it has been speculated that the most powerful tool for such an
analysis would be the cross-correlations between different galaxy
populations~\cite{McDonald:2009ud,Yoo:2012se,croft13,Bonvin:2013ogt,Alonso:2015sfa,Fonseca:2015laa}. The
correlation of two differently biased tracers of the underlying matter
field will, at least to the first order, boost the signal of the beyond-Kaiser
effects relative to the difference in the density bias factors. Thus
the larger the difference the stronger the signal. Unfortunately the
difference in density bias factors among different galaxy populations
is relatively small, and the idea has only received attention very recently.

On the other hand, the tracers such as \lya\ forest or QSOs that are
measured in the same survey volume by already existing surveys are
very differently biased tracers. Exploiting this difference would
boost the signal of the relativistic effects to a measurable
level.
Given the highly non-linear relation between the observed
\lya\ forest flux fluctuations and the underlying density field, one
might question the ability to use the \lya\ forest as a linear
tracer. However, several studies have shown that on large scales
\lya\ forest behaves as any other local tracer and any deviations
start to show up at smaller scales~\cite{arinyo15,slosar11,mcdonald03}. 

In this paper we compute the theoretical predictions for the
QSO-\lya\ cross-correlation function and show how the difference in
the density biases of the two tracers boosts the signal of the
relativistic effects to a 10\% level above scales of
$\sim40\;h^{-1}\mathrm{Mpc}$. 
The paper is structured as
follows: in Sections~\ref{sec:LyaF}  and \ref{sec:LyaO} we introduce
the \lya\ forest observable and derive the flux fluctuations using
fully relativistic treatment. In Section~\ref{sec:xi} we apply the
calculations to the QSO-\lya\ cross-correlation function and present
the results for both symmetric and anti-symmetric parts. We conclude in
Section~\ref{sec:conc}.

\section{Ly-$\alpha$ absorption}
\label{sec:LyaF}

The \lya\ forest is an absorption feature seen in the spectra
of high redshift QSO and galaxies and is a unique tracer of
the underlying matter distribution in the redshift range of $z=2-5$.

When the light from a distant QSO
passes through the intergalactic medium (IGM), the neutral hydrogen along the line of sight will
cause absorption of the QSO continuum by the redshifted \lya\ 
($121.565\;\mathrm{nm}$) resonance line. The collection of this absorption lines
blend into one feature called the \lya\ forest. In fact it is more correct
to think of the \lya\ forest as being caused by continuous absorption
of the mildly fluctuating density of the IGM.

Observed transmission of the flux is thus a tracer of the baryon
distribution of the IGM and one of the successes of the \lya\ in the
last few years has been to establish it as a precision cosmology
probe. 

The photons' absorption due to \lya\ is described by the differential equation
\be  \label{photon_abs}
d \ln \NN_\gamma = -  A\ n_{\text{HI}} \  \phi\! \left( E, E_\alpha \right) dt 
\ee
where $E_\alpha$ is the Lyman-$\alpha$ absorption line, $
\mathcal{N}_\gamma$ the photon occupation number, $t$ the is the
proper time in absorption HI rest frame and $n_\text{HI}$ the number
density of neutral hydrogen. $A$ is a physical constant derived from Einstein
coefficients and given by $A = e^2 f_\alpha/4\varepsilon_0 m_e$, 
where $e$ is the electron charge, $\varepsilon_0$ is vacuum
permittivity, $m_e$ electron mass and $f_\alpha=0.4164$ oscillation strength
of the \lya\ transition. The integral of the absorption line profile ($\phi\! \left(
E, E_\alpha \right)$) over the energy is normalized to the unity. 
We define the optical depth as
\be
\label{tau_def}
\tau  \equiv  \int_\text{light-cone} d \ln \NN_\gamma.
\ee
The above definition agrees with the non-relativistic result in the
appropriate limit.

\section{Ly-$\alpha$ observable}
\label{sec:LyaO}
On large scales we can work under the approximation that $\phi \left(
E, E_\alpha \right) = \delta_D \left( E - E_\alpha \right)$. Hence, we
neglect thermal broadening which affects scales of
$10\; h^{-1} \mathrm{\;Mpc}$. This approximation is valid as long as we
are interested in large scale correlations.
Therefore, our observable is the transmitted flux fraction $F$, 
i.e.~the photon occupation number observed normalized with the emitted one, which yields to
\be \label{Flux_def}
F=e^{- \tau } \, ,
\ee
in terms of the measured angle\footnote{We follow the notation of Ref.~\cite{Bonvin:2011bg}, where $\bn$ denotes the direction of propagation of the photon. Hence the measured angle is $- \bn$, but for sake of simplicity will refer to $\bn$ as argument of the observable quantities.} $\bn$ and redshift $z$, corresponding to the shift in the Lyman-$\alpha$ spectrum (i.e.~$1+z= E_\alpha /E$), namely 
$F\left(\bn , z \right)$. One can define the fluctuations of the
transmitted flux as
\be
\delta_F \left( \bn , z \right) \equiv \frac{F \left( \bn , z \right) }{ \langle F \left( \bn , z \right) \rangle } -1 \, ,
\ee  
where $\langle . \rangle $ denotes the angular mean at fixed observed redshift $z$.

On large scales, we can linearize the relation between the transmitted
flux and the optical depth and express the observable quantity as
\be \label{DeltaF_linear}
\delta_F \left( \bn , z \right) = - \left( \tau \left( \bn , z \right) - \langle \tau \left( \bn , z \right) \rangle \right) \equiv - \delta \tau \left( \bn , z \right) \, .
\ee

The linearization of the exponential in Eq.~(\ref{Flux_def}) might
question the validity of the perturbative approach. Nevertheless, as
shown in section~\ref{sec:biases}, one can expand perturbatively the
transmitted flux $F$ in terms of the optical depth $\tau$ by
introducing some bias factors which encode the effect of the mode
coupling.

From Eqs.~(\ref{photon_abs}) and (\ref{tau_def}) we obtain 
\bea
\label{dlnN}
\tau \left( \bn , z \right)   &=& - A \int n_{\text{HI}} \ \delta_D  \! \left( E  - E_\alpha \right)   dt  = - A \int n_\text{HI} \delta_D \left( E^\text{obs} \left( 1 + z \right) - E_\alpha \right) dt 
\nonumber 
\\
&=& - A \int \frac{n_\text{HI}}{E^\text{obs}} \delta_D \left(  \left( 1 + z \right) - \frac{E_\alpha}{E^\text{obs}} \right) \frac{dt}{dz}dz 
= - A n_\text{HI} \frac{ 1 + z }{E_\alpha} \frac{dt}{dz}
\eea
where all quantities are evaluated at $1+z = E_\alpha/E_{\text{obs}}$ and the suffix 'obs' denotes quantities evaluated with respect to the observer rest frame and $t$ is the proper time\footnote{We prefer not to denote the proper time with $\tau$ to avoid confusion with the optical depth.}  in the photons absorption frame.

 In this section we aim to derive Eq.~(\ref{DeltaF_linear}) to first order in perturbation theory. As long as we describe observable quantities, we have the freedom to work in any gauge without loss of generality. We choose to work in Newtonian gauge $d\tilde s ^2 = a^2 ds^2$, where $a$ is the scale factor of the universe and
\be
d s^2 =  - \left( 1+ 2 \Psi \right) d \eta^2 + \left( 1 - 2 \Phi \right) \delta_{ij}dx^i dx^j 
\ee
is the conformal metric, where the metric perturbations $\Psi$ and $\Phi$ are the Bardeen potentials. From Eq.~(\ref{dlnN}), we need to compute $\frac{dt}{dz}$ to first order in perturbation theory
\bea
\label{dtdz}
\frac{dt}{dz} &=& \frac{dt}{d\eta} \frac{d\eta}{dz}  = \frac{dt}{d\eta} \frac{d\eta}{dz} = 
\frac{1+ \Psi}{1+ \bar z} \left( \frac{d\bar z}{d \eta} + \frac{d \delta z}{d\eta} \right)^{-1} 
\nonumber \\
&=&
 \frac{1+ \Psi}{1+ \bar z} \left( \frac{d\bar z}{d \eta} \left( 1 + \frac{\delta z}{1+z}  \right)+ \left( 1 + z \right) \frac{d}{d\eta} \left( \frac{\delta z}{1 +z } \right) \right)^{-1} 
\nonumber \\
&=&
- \frac{1 + \Psi}{\HH \left( \bar z \right) \left( 1 + \bar z \right)^2} \left[ 1 - \frac{\delta z}{1+ z} + \frac{1}{\HH} \frac{d}{d\eta} \left( \frac{\delta z}{1 +z } \right) \right] 
\nonumber \\
&=&
- \frac{1}{\HH \left( 1 + z \right)^2} \left[ 1 + \Psi + \left( 1 - \frac{\dot \HH}{\HH^2} \right)\frac{\delta z}{1+z} + \frac{1}{\HH} \frac{d}{d\eta} \left( \frac{\delta z}{1 +z } \right) \right] 
\nonumber \\
&=& 
- \frac{1}{\HH \left( 1 + z \right)^2}  
\left[ 1 + \Psi  + \left( 1 - \frac{\dot \HH}{\HH^2} \right)\frac{\delta z}{1+z} +\ndv + \HH^{-1} \partial_r \ndv + \HH^{-1}\dot \Phi \right] \, , \ \
\eea
where $\HH= \dot a/a$ is the comoving Hubble parameter and a dot denotes the derivative with respect to the conformal time $\eta$. We have also introduced the unperturbed redshift $\bar z$ by expanding the observed redshift as $z= \bar z+ \delta z$. In general, in this work, we will denote with an overbar background quantities in perturbation theory.
In the last equality we have assumed that sources move along geodesic
\be
\bn \cdot \dot \bv + \HH \ndv - \partial_r \Psi = 0 \, ,
\ee
and used
\be
 \frac{\delta z}{1+z}  = - \left(   \Psi + \ndv +  \int_\eta^{\eta_o} \left( \dot \Psi + \dot \Phi \right) d\eta'  \right) \, ,
\ee
which is derived from geodesic equation.
We remind that the physical density of HI, i.e.~$n_\text{HI} \left( z \right) $ in Eq.~(\ref{dlnN}), is related to the density computed in perturbation theory to first order as
\be
n_\text{HI} \left( z \right) =\bar n_\text{HI} \left( z \right) \left( 1 - \frac{d \ln {\bar n}_\text{HI}}{d z} \delta z + \delta_\text{HI} \right) \, .
\ee
The second term in the brackets describes the source evolution and it is often parametrized through the evolution bias
\be \label{fevo}
f_\text{evo}^{\elya} \equiv \frac{\partial \ln \left( a^3 \bar n_\text{HI} \right)}{\HH \partial \eta}
= 3 - \left( 1+ z \right) \frac{d \ln \bar n_\text{HI} } {dz} \, .
\ee
Finally, we find
\bea \label{tau_fin}
\tau \left( \bn ,z \right) &=& {\bar \tau} \left( z \right)
\left[ 1 + \delta^\text{sync}_\text{HI} + \HH^{-1} \partial_r \ndv  - \left( 2 + \frac{\dot \HH}{\HH^2} - f_\text{evo}^{\elya}  \right)\frac{\delta z}{1+z} 
\right.
\nonumber \\
&&
\left.
\qquad \qquad
+\ndv + \Psi   + \HH^{-1}\dot \Phi  +  \left(  f_\text{evo}^{\elya} -3  \right) \HH v \right] \, .
\eea
where 
\be
{\bar \tau} \left( z \right) \equiv  \frac{A}{E_\alpha} \frac{\bar n_\text{HI} }{\HH \left( 1 + z \right)} 
\ee
and $v$ is the (gauge invariant) velocity potential in Newtonian gauge defined through $\bv = - {\bf \nabla} v$. We have also expressed the density perturbation in synchronous gauge, through $\delta^\text{sync} = \delta + \left( 3 - f_\text{evo} \right) \HH v$.

The unperturbed optical depth ${\bar \tau}$ can be rewritten as (\cite{mcdonald03}):
\bea
\label{tau0}
{\bar \tau} \left( z \right) &=& \frac{A {\bar n}_\text{HI} }{E_\alpha H } \approx 1.27 \times \frac{10^{-12} s^{-1} }{\Gamma_{\gamma,\text{HI}}} \left( \frac{X}{0.75} \right)^2 \left( \frac{T_0}{10^4 \text{K} } \right)^{-0.7} \frac{\omega_{b,0}^2}{h} \frac{H_0}{H} \left( 1+ z\right)^6,
\eea
where $T_0$ is the mean temperature of the gas
and we have used the mass fraction of neutral hydrogen through the
photo-ionization equilibrium
\be
n_\text{HI} = n_{H}^2 \frac{\alpha \left( T \right)}{\Gamma_{\gamma,\text{HI}}}  = \frac{X^2 \rho_b^2 }{m_\text{H}^2}\frac{\alpha \left( T \right)}{\Gamma_{\gamma,\text{HI}}}  \, .
\ee
where $\Gamma_{\gamma,\text{HI}}$ is the photo-ionization rate, $X$ is
the hydrogen mass fraction, $n_{H}$ is the hydrogen number density,
$m_{H}$ is the hydrogen mass, $\rho_b$ is the baryon energy density
and the recombination rate $\alpha(T)$ is given by
\be
\alpha \left( T \right) = \alpha_0 \left( \frac{T}{10^4 \text{K} } \right)^{-0.7} \, ,
\ee
where $\alpha_0 = 4.3\,\times 10^{-13}\;\mathrm{s^{-1}\,cm^3}$.

In this work we consider the Planck best fit cosmological parameters~\cite{Ade:2015xua}, namely $\omega_{b,0} = 0.022$, $h=0.67$ and the astrophysical parameters $T_0= 1.47088 \times 10^4\; \text{K}$, $X=0.75$ and $\Gamma_{\gamma,\text{HI}} = 10^{-12} s^{-1}$. With these parameters Eq.~(\ref{tau0}) reduces to
\be
{\bar \tau}\left( z \right) \approx 7 \times 10^{-4} \frac{\left( 1+z \right)^6}{E\left( z \right)}
\ee
with $E(z) = H(z)/H_0$.

\subsection{Quasar density fluctuations}

In the following of this paper we will be interested in
cross-correlating the \lya\ signal with the quasar density
fluctuations. In the light of that we give an expression with
relativistic corrections used for the QSO fluctuations as found in the
literature, and used in this paper.

The number counts for discrete tracers has been derived in~\cite{Challinor:2011bk,Bonvin:2011bg} in galaxy clustering framework. Nevertheless their results can be applied as well to QSOs, just considering the appropriate bias factors, 
\bea \label{delta_Q}
 \Delta_Q \left( \bn, z \right) &=& b_Q \delta^\text{sync} +(5s -2)\Phi + \Psi + \frac{1}{\HH}
\left[\dot \Phi+\dd_r(\ndv)\right] +  \left( f^Q_\text{evo} - 3 \right) \HH v\nonumber \\  &&
+ \left(\frac{{\dot\HH}}{\HH^2}+\frac{2-5s}{r_S\HH} +5s-f^Q_{\rm evo}\right)\left(\Psi+\ndv+ 
 \int_0^{r_S}\hspace{-0.3mm}dr(\dot \Phi+\dot \Psi)\right) 
   \nonumber \\  &&  \label{DezNF}
+\frac{2-5s}{2r_S}\int_0^{r_S}\hspace{-0.3mm}dr \left[2-\frac{r_S-r}{r}\Delta_\Om\right] (\Phi+\Psi) \,.
\eea
Differently from \lya\ forest, single tracer number counts include a cosmic magnification contribution and a magnification bias parameter $s$ defined as
\be
s= - \left. \frac{2}{5} \frac{\partial \ln \bar n \left( z, \ln L \right)}{\partial \ln L} \right|_{\bar L}
\ee
where $\bar L$ denotes the threshold luminosity of the survey and $\bar
n$ is the background number density.
Similarly, the evolution bias factor of QSOs is related to the QSO
number density distribution (see Eq.~(\ref{fevo})).

\subsection{Non-linear corrections}
\label{sec:biases}

In the previous section we have computed the optical depth $\tau(z)$
to first order in perturbation theory. However, the exponential
(\ref{Flux_def}) introduces a mode coupling that we neglect by
linearizing this expression. In the context of the relativistic
corrections beyond simple Kaiser approximation this is valid if the
terms themselves are small. 

In this section we give an estimation of the bias for the relativistic
terms (mainly Doppler term as it is largest). We use the formalism
described in~\cite{seljak12,cieplak15}, and extend it to capture
relativistic correction described in this paper. We do not use this
linear description to try to describe the density bias.

If the terms are not small, a correction capturing the
non-linearity of the $F-\tau$ transformation is applied in terms of
bias factors, as is the case with the redshift-space distortions. The
full calculation for the velocity-gradient bias factor can be found in
the literature (\cite{seljak12,cieplak15}), but we will review the
basic steps here for clarity. To the leading order, the relativistic
corrections enter our \lya\ calculations only through the change of
coordinates, from the real space time (radial coordinate) to a
redshift (redshift-space coordinate) as can be seen from
Eqs.~\ref{dlnN}-\ref{dtdz}. Neglecting for the moment the relativistic
corrections and focusing only on the velocity-gradient part, we see
that the observed optical depth can be written as
\be
\label{tau_bias_rsd}
\tau(\bn,z) = {\bar \tau}\left(1+\delta\right)^p\left( 1 +
\HH^{-1}\partial_r \ndv \right),
\ee
where the value of $p$ is usually taken in the range $\sim 1.5-1.7$ \cite{seljak12},
and $\delta$
is the non-linear density fluctuations.
From here one can define an optical depth bias with respect to the
redshift-space distortions as
\be
b_{\tau,v} \equiv \frac{1}{\langle \tau\rangle} \left\langle \frac{\partial
  \tau}{\partial\left(\HH^{-1} \partial_r \ndv \right)} \right\rangle = 1.
\ee
This is not a new result, indeed it tells us that there is no
velocity-gradient bias, as one would expect. However, the observable
of the \lya\ forest is not the optical depth but the flux, and the
non-linear transformation is applied after the redshift-space
distortions. One can
write the flux bias of the velocity-gradient as
\bea
b_{F,v} &\equiv& \frac{1}{\langle F\rangle} \left\langle \frac{\partial
  F}{\partial\left(\HH^{-1} \partial_r \ndv \right)} \right\rangle =
\frac{1}{\langle F\rangle} \left\langle \frac{\partial
  F}{\partial\left(\HH^{-1} \partial_r \ndv \right)} \right\rangle = \notag \\
&=&
\frac{1}{\langle F\rangle} \left\langle \frac{\partial F}{\partial
\tau} \frac{\partial \tau}{\partial\left(\HH^{-1} \partial_r \ndv
  \right)} \right\rangle = \frac{1}{\langle F\rangle} \left\langle \frac{\partial F}{\partial
\tau} \tau \right\rangle = \notag \\ 
&=& \frac{\langle F \ln{F} \rangle}{\langle F\rangle}.
\eea
This result has been first obtained by~\cite{seljak12} and has been
further shown to work extremely well in the absence of thermal
smoothing (\cite{cieplak15}). 

The above review calculation was done using
redshift-space distortions only, as written in
Eq.~\ref{tau_bias_rsd}. However there is no reason the same
calculation can be carried out in the same manner if one replaces
$\HH^{-1}\partial_r \ndv$ with any other quantity, e.g. a relativistic
doppler term ($\ndv$). In such a case one would modify the optical
depth relation as
\be
\label{tau_bias_dop}
\tau(\bn,z) = {\bar \tau}\left(1+\delta\right)^p\left( 1 +
\HH^{-1}\partial_r \ndv + \left(3 - f_{\text{evo}}^{\elya} + \frac{\dot
  \HH}{\HH^2}\right) \ndv \right).
\ee
Accordingly the optical depth bias, with respect to Doppler term would
now be
\be
b_{\tau,D} \equiv \frac{1}{\langle \tau\rangle} \left\langle \frac{\partial
  \tau}{\partial \left(\ndv \right)} \right\rangle  = 3 - f_{\text{evo}}^{\elya} + \frac{\dot
  \HH}{\HH^2},
\ee
and since the $b_{\tau,D}$ is a constant it furthermore means
\be \label{Doppler_bias}
b_{F,D} = b_{\tau,D} \frac{\langle F \ln{F} \rangle}{\langle F\rangle}
= b_{\tau,D} b_{F,v}.
\ee

Thus any kind of scaling of the redshift coordinate that is linearly
added to the redshift space distortions acquires the same (flux) bias factor due to
non-linear transformation between the flux and the optical depth, with
a pre-factor that is nothing else but the optical depth bias factor
for the same quantity.

The above calculation was carried out in the limit of no thermal
broadening, since our paper focuses only on the large scale
limit. However there is no reason that thermal broadening would
distinguish between velocity-gradient effect or Doppler term effect,
thus resulting in the same correction to the overall bias factor.

Moreover, in the above model no second order effects were included. 
In that sense it should be tested on numerical simulations, 
and only draws on the fact that the leading
relativistic effects considered in this paper are nothing less than
small corrections to redshift-space transformation, compared to the
redshift space distortion. Thus they would behave in the same way
under the non-linear transformation. To fully understand the values of
the bias factors of the \lya\ forest a more detailed analysis is
required that is beyond the scope of this paper.


Therefore to correctly describe the transmitted flux fluctuation we
change the bias factors predicted by the linear
theory with the observed bias factors
\bea \label{Lya_Delta}
\delta_F \left( \bn , z \right) &=&  b_\alpha \delta^\text{sync} + b_v
\HH^{-1} \partial_r \ndv  + b_R \!
\left[ - \!\left( \!2 + \frac{\dot \HH}{\HH^2} - f_\text{evo}^{\elya} \!\right)\frac{\delta z}{1+z} \right.
\nonumber \\
&& 
\hspace{2cm}
\left.
+\ndv + \Psi   + \HH^{-1}\dot \Phi  + \left( f_\text{evo}^{\elya}-  3 \right)  \HH v \!\right] \, .
\eea

The first two terms correspond to a simple Kaiser approximation for the \lya\ forest tracer, while the terms in the square bracket describe the relativistic corrections.
Note that the
velocity-gradient bias different than unity is the result of the redshift space
distortion acting on the optical depth, before taking the non-linear
transformation to the observable which is the flux, and it is not to
be interpreted as a violation of the equivalence principle. 
This result is known and widely used in the
literature~\cite{mcdonald03,seljak12,slosar11,arinyo15}.
According to the model presented in this section we adopt a universal
bias for all relativistic terms, $b_R$, which is to the correct
description to linear order.

\subsection{Biases}

In a relativistic framework one has to define the bias parameters with
respect the density, and redshift space distortions for \lya\ forest
tracer, in a specific gauge. Indeed, if on small scales the
differences between different gauges are generally suppressed by a
factor $\left(\HH/k\right)^2$, on large scales the gauge choice
becomes relevant and it might introduce some spurious $k$-dependences
in the bias factors.

However they do not affect the main
results of our work which is determined by terms suppressed only by
$\HH/k$. Our choice to apply density bias $b_\alpha$ in
synchronous comoving gauge is well justified from the assumption
that the tracer and the underlying density perturbation
experience the same gravitational field and they move with the same
velocity. Hence in their rest frame we can apply the linear bias
prescription. In the standard approach the velocity-gradient bias is
applied to the Kaiser term, and is derived using Newtonian relation
between time derivative of density perturbation and peculiar
velocity. If Cold Dark Matter (CDM) is the only species which
contributes to energy-momentum tensor perturbations, this relation is
equivalent to
the linearized Einstein equation provided that the density
perturbation is expressed in the synchronous and velocity in
the newtonian gauge. Therefore, to be consistent with standard analysis we
used velocity-gradient bias in newtonian gauge.

In this work we consider the \lya\ density bias evolution as suggested
by the observations (\cite{mcdonald06,slosar11,slosar13,busca13})
\be 
b_\alpha(z) = b_\alpha(z=2.5) \left(\frac{1+z}{3.5}\right)^{2.9}
\ee
with $b_\alpha(z=2.5) = -0.15$. For the velocity-gradient bias of the
\lya\ forest we follow the recent work using simulations (\cite{arinyo15}) which gives:
\be \label{vel_bias_def}
 b_v \equiv \frac{\beta_\alpha(z) b_\alpha(z)}{f(z)}, \qquad
\beta_\alpha(z) = \beta_\alpha(z=2.5) \left(\frac{1+z}{3.5}\right)^{-0.86},
\ee
with the value at the mean redshift $\beta_\alpha(z=2.5) = 1.36$, and
$f=d \ln \delta / d\ln a$ is the growth rate factor. 

In addition we have a non-vanishing evolution factor for \lya\ forest,
i.e.~$f_\text{evo}^{\elya} \neq 0$ in Eq.~(\ref{Lya_Delta}). Following
Eq.~(\ref{fevo}) the value of the \lya\ evolution bias can be
predicted from the redshift evolution of the neutral hydrogen
fraction. Using the result of Eq.~(\ref{tau0}), we see that 
$n_{HI}(z) \sim \Gamma_{\gamma,HI}^{-1} T_0^{-0.7} \rho_b^2$. For
${\bar \tau}$, we have followed the calculations of
(\cite{mcdonald03}), and assumed that both UV background radiation
($\Gamma_{\gamma,HI}$) and the mean temperature of the gas $T_0$ are
constant with redshift, thus resulting in a simple evolution of
neutral hydrogen $n_{HI}\sim\left(1+z\right)^6$. In turn this implies
a value of $f_{evo}^{\elya} = -3$. This is the value used in the
calculations in the remainder of this paper, unless stated otherwise.

Moreover, for the \lya\ forest, we also have bias of the relativistic
terms. Given the definition of such a bias (Eq.~\ref{Doppler_bias}
and~\ref{Lya_Delta}) we use an approximation for its value and say
\be
b_R(z) = b_v(z).
\ee

For the QSO density bias factor ($b_Q$) we use the semi-empirical
relation derived by~\cite{croom05,myers07}:
\be
b_Q(z) = 0.53 + 0.289\left( 1+z\right)^2
\ee
The empirical relation fits the observed data very well at lower redshifts
and can be safely extrapolated to our mean redshift of $z=2.5$, giving
a value of $b_Q(z=2.5) \sim 4$, which is a bit high, but in agreement with independent
observations at higher redshifts~\cite{white12,font13,font14}.

To derive both the evolution ($f_{\text{evo}}^Q$) and magnification
($s$) bias parameters we have used a fitting model for Quasar Luminosity Function
used in BOSS DR9 data analysis~\cite{ross12QSOLF}. The assumed
threshold absolute magnitude was $\bar M = -24.5$. The derived values
for the magnification and evolution bias factors are $s = 0.295319$
and $f_{\text{evo}}^Q = 5.7999$. 

\section{2-point correlation function}
\label{sec:xi}

Having derived the \lya\ forest observable and defined the QSO
observable we are now interested in studying the 2-point function of the correlation between \lya\ and QSOs. Namely 
\be \label{2point}
\xi_{Q\alpha} \left( z_1 , z_2 , \theta \right) \equiv \langle \Delta_Q \left( \bn_1, z_1 \right) \delta_F \left( \bn_2 , z_2 \right) \rangle
\ee
where $\cos \theta = \bn_1 \cdot \bn_2$ and  $\Delta_Q \left( \bn, z
\right)$ is the QSO number counts. Equivalently one can define the
cross-correlation where the positions of Ly-$\alpha$ and QSO have been
exchanged:
\be
\label{2point_aQ}
\xi_{\alpha Q} \left( z_1 , z_2 , \theta \right) \equiv \langle \delta_F \left( \bn_1, z_1 \right) \Delta_Q \left( \bn_2 , z_2 \right) \rangle.
\ee

 \begin{figure}[!h]
  \centering
  \includegraphics[width=0.4\linewidth]{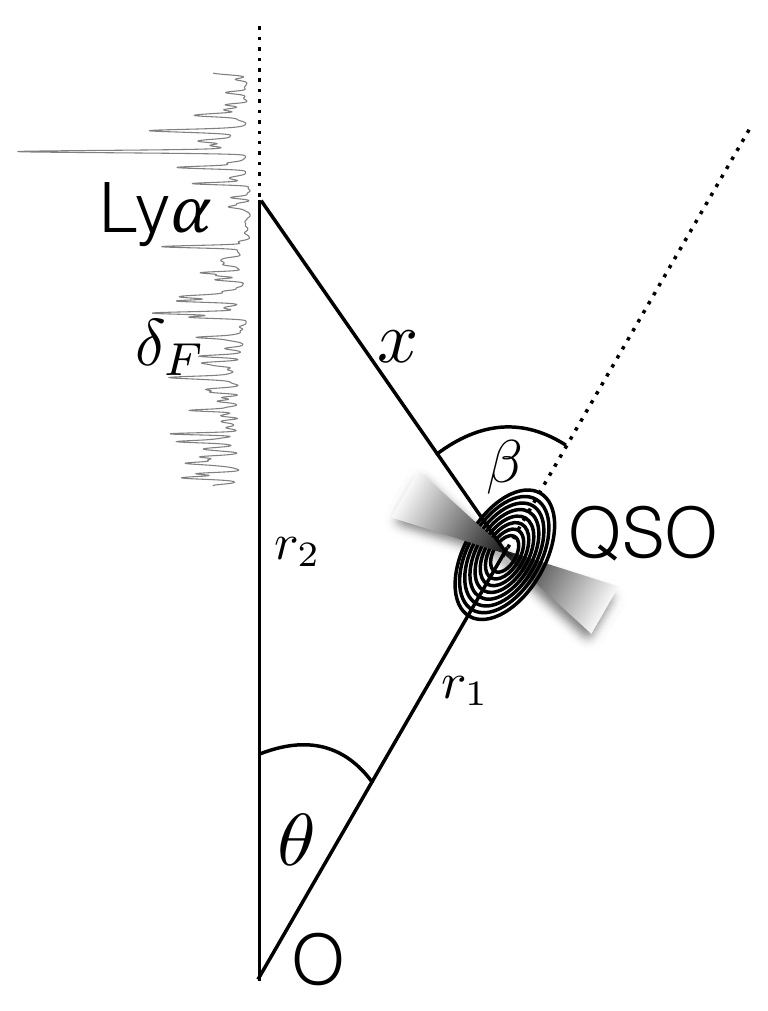}
  \caption{ \label{cartoon} The figure illustrates the correlation between Quasars and \lya\ forest. The comoving distances of the sources from the observer are denoted by $r_1$ and $r_2$, while $x$ refers to the comoving separation between them.
  }
\end{figure}

From Fig.~\ref{cartoon} we remark that, by assuming a fiducial cosmology, we can relate the angles $\beta$ and $\theta$. Hence we can express the correlation functions~(\ref{2point}, \ref{2point_aQ}) in terms of the triplet $\left( z_1, z_2 , \beta \right)$, and we denote them as $\hat\xi_{\alpha Q}$ and $\hat\xi_{\alpha Q}$.
We see direclty that the transformation of
coordinates ($z_1 \leftrightarrow z_2$, $\beta \rightarrow \pi-\beta$)
maps from one to the other cross-correlation functions,
$\hat\xi_{Q \alpha}(z_2,z_1,\pi-\beta) = \hat\xi_{\alpha
  Q}(z_1,z_2,\beta)$. While the interchange of the redshifts is
obvious, one can see from Fig.~\ref{cartoon} that the angle that needs
to be flipped is $\beta$, the angle between the QSO and the pixel in
\lya\ forest, while the angle $\theta$ describes their angular
separation. The angles are of course connected and either can be used
to fully describe the cross-correlation function, however $\theta$ is
a more natural choice to use with the angular power spectrum expansion
(see below).

 In general, it is always possible to express the 2-point correlation function in terms of angular power spectra $C_\ell \left( z_1 ,z_2 \right)$ as
\be \label{2point_cl}
\xi\left( z_1 , z_2 , \theta \right) = \frac{1}{4 \pi} \sum_\ell \left( 2 \ell + 1 \right) C_\ell \left( z_1 , z_2 \right) P_\ell \left( \cos \theta  \right) \, ,
\ee
where $P_\ell$ denote the Legendre polynomials. In appendix~\ref{sec:cl} we introduce the corresponding angular power spectra. Since we are mainly interested in the 1-dimensional correlation function along the radial direction a direct evaluation of Eq.~(\ref{2point}) is simpler and more intuitive than passing through Eq.~(\ref{2point_cl}). Nevertheless, we prefer to use Eq.~(\ref{2point_cl}) to handle the integrated terms (cosmic magnification, integrated Sachs-Wolfe, Shapiro time-delay) and take advantages of \class{} code~\cite{Lesgourgues:2011re,Blas:2011rf}, whose modified transfer functions are shown in appendix~\ref{sec:CLASS}, than rely on approximations.

For the 1-dimensional radial correlation, Eq.~(\ref{2point}) reduces to a function of two redshifts $z_1$ and $z_2$. It is then convenient to re-expressed them as a redshift separation between the two source positions $\Delta z= z_1 - z_2$ and a mean redshift $z_\text{mean} = \left( z_1 + z_2 \right)/2$. By assuming a fiducial model we can always map the redshift separation in terms of comoving separation as $x\left( z_1 , z_2 \right) \equiv r\left( z_1\right) - r \left( z_2 \right)$, where 
\be \label{com_dist}
r(z) =  \int_0^z\frac{dz'}{H(z')}
\ee
is the unperturbed comoving distance and
\be
H^2(z) = H_0^2\left(a^{-3}\Om_m+ a^{-2}\Om_K +\Om_\La\right) \, .
\ee
We remark that to first order in perturbation theory it is enough to
use the background unperturbed redshift-distance relation
Eq.~(\ref{com_dist}). Because, in general, sources are not
homogeneously distributed along the redshift, the 1-dimensional
correlation function will not depend only on the absolute value of
$\Delta z$, or $x\left(z_1, z_2 \right)$, but on its sign as
well. 

This change of coordinates simplifies the relation when switching the
places of \lya\ and QSO, $\xi_{\alpha Q}(x, z_{\text{mean}})$ and
$\xi_{Q \alpha}(x,z_{\text{mean}})$, since the only transformation required is 
$x\rightarrow -x$.

Ideally, since the pure radial correlation is described by a single line of sight, the light emitted by a QSO at high redshift would be partially absorbed by the IGM at lower redshift and it will generate the \lya\ forest absorption lines. This situation is clearly a violation of the photon number conservation assumption on which the derivation (see for instance Ref.~\cite{Challinor:2011bk,Bonvin:2011bg}) of number counts observable for discrete tracers is based. Nevertheless in our work we never consider this situation and we assume that the two tracers, \lya\ forest and QSO are correlated gravitationally only. 
This assumption is consistent with what is done in observations, where the
correlation between the \lya\ forest and the QSO, which produces the
absorption lines, is not considered.
Another issue is due to the fact that light emitted by a QSO at
high redshift would be affected also by the gravitational potential
generated by the \lya\ forest through the cosmic magnification
effect. Being the \lya\ forest and the QSO on the same line of
sight, the impact parameter of the light deflection vanishes and we
can not anymore considered it in the weak lensing regime. Since, in any
case, exact radial correlation are not taken into consideration we
neglect this effect and we describe it as a weak lensing
phenomenon. Let us also stress that these possible issues are
completely relieved once we consider the dipole with respect to the
angle $\beta$ defined in Fig.~\ref{cartoon}. Indeed
by averaging over all angles the contribution of one single direction
has not weight in the integral and its contribution should be
negligible. We leave the analysis of the dipole of the correlation function to a future work.

We compute then the full relativistic 2-point correlation function 
\be
\xi_{Q\alpha} = \xi_{Q\alpha}^\text{newt} +   \xi_{Q\alpha}^\text{magnification} + \xi_{Q\alpha}^\text{relativistic}
\ee
where we dropped the dependence on $\Delta z$ and $z_\text{mean}$ for simplicity. We start by introducing the primordial curvature power spectrum through
\be
\langle R_\text{in} \left( \bk \right) R_\text{in} \left( \bk' \right)   \rangle = \left( 2 \pi \right)^3 \delta_D^{(3)} \left( \bk + \bk' \right) P_R \left( k \right) 
\ee
and we write any first order variable $X$ in terms of transfer functions normalized with the primordial curvature as
\be
X \left( \bk, \eta \right) = T_X \left( k, \eta \right) R_\text{in} \left( \bk \right)  \, .
\ee

Then, by computing the different contributions we find 
\bea 
\label{xi_newt}
\xi_{Q \alpha}^\text{newt} &\equiv& \langle \left[ b_Q \delta+ \HH^{-1} \dd_r(\ndv)\right] \left( \bn, z_1 \right) \left[b_\alpha \delta+ b_v \HH^{-1} \dd_r(\ndv) \right]\left( \bn , z_2 \right) \rangle 
\nonumber \\
&=& b_Q \left( z_1\right)  b_\alpha \left( z_2 \right) \int \frac{d k }{2 \pi^2} k^2 T_\delta \left( k , \eta_1 \right)  T_\delta \left( k , \eta_2 \right) P_R \left( k \right) j_0 \left(   k x\right)   
\nonumber \\
&&+ b_v \left( z_2 \right) \int \frac{dk }{2 \pi^2} \frac{k^4}{\HH\left( z_1\right) \HH\left( z_2 \right)} T_v  \left( k , \eta_1 \right)  T_v \left( k , \eta_2 \right) P_R \left( k \right)
\nonumber \\
&& \qquad \qquad
 \times \left[ \frac{1}{5} j_0  \left(   k x \right)  - \frac{4}{7} j_2 \left(   k x \right) + \frac{8}{35} j_4 \left(   k x \right)\right]
 \nonumber \\
 && +
 \int \frac{dk}{2 \pi^2} k^3   P_R \left( k \right)  \left[ \frac{  b_Q \left( z_1 \right) b_v \left( z_2 \right)}{\HH \left( z_2 \right) } T_\delta \left( k , \eta_1 \right) T_v \left( k , \eta_2 \right)     
 + \frac{   b_\alpha \left( z_2 \right)}{\HH \left( z_1 \right) } T_v \left( k , \eta_1 \right) T_\delta \left( k , \eta_2 \right)
  \right]
  \nonumber \\
  && \qquad \qquad
 \times \left[- \frac{1}{3} j_0 \left(   k x \right)  + \frac{2}{3} j_2 \left(   k x \right)  \right]
 \eea
where $x= x \left( z_1 , z_2 \right) $,  and we allow for redshift dependent bias factors, while we have implicitly assumed that these are scale independent. Nevertheless a generalization to scale dependent bias factor would be straightforward. We define the correlation function induced by cosmic magnification as follows
\be
\xi_{Q \alpha}^\text{magnification} \equiv \langle \left[ \frac{5s-2}{2r}\int_0^{r}\!\!\!\!dr' \left[\frac{r-r'}{r'}\Delta_\Om\right] (\Phi \!+\!\Psi)  \right] \!\!\left( \bn , z_1 \right)  \left[b_\alpha \delta+ b_v \HH^{-1} \dd_r(\ndv) \right]\!\!\left( \bn , z_2 \right)  \rangle \, .
\ee
Because of the integral along the line of sight this expression can not written in a simple form like Eq.~(\ref{xi_newt}), unless we adopt Limber approximation~\cite{LoVerde:2008re}. We therefore prefer to compute explicitly this contribution through Eq.~(\ref{2point_cl}).
We define the relativistic contribution as follows
\bea
\label{xi_rel}
\xi_{Q \alpha}^\text{relativistic} &\equiv& \langle \Delta_Q \left( \bn, z_1 \right) \delta_F \left( \bn , z_2 \right) \rangle - \xi_{Q \alpha}^\text{newt} - \xi_{Q \alpha}^\text{magnification} 
\nonumber  \\
& \equiv& \xi_{Q \alpha}^{ \text{Doppler}} + \xi_{Q \alpha}^{\text{potential}} + \xi_{Q \alpha}^\text{non-local}
\eea

where we have a Doppler term\footnote{For the velocity transfer function $T_v \left( k , \eta \right)$  we follow the notation of Ref.~\cite{Bonvin:2011bg,DiDio:2013bqa}, namely it refers to vector potential $V$ defined through $\bv = i \hat \bk V$.}
\bea
\xi_{Q \alpha}^{ \text{Doppler}} &\equiv& \langle \left[ 
[ b_Q \delta+ \HH^{-1} \dd_r(\ndv) + {\mathcal{R}_Q}  \ndv
\right] \left( \bn, z_1 \right)
 \left[b_\alpha \delta+ b_v \HH^{-1} \dd_r(\ndv)  + {\mathcal{R}_\alpha} \ndv \right]\left( \bn , z_2 \right) 
\rangle
\nonumber \\
&&
 -  \xi_{Q \alpha}^\text{newt}
\nonumber
\\
&=& \int \frac{dk}{2 \pi^2} k^2  P_R \left( k \right)
 j_1 \left( k x \right) 
 \nonumber \\
 && \qquad
 \left[ 
 b_Q \left( z_1 \right) {\mathcal{R}_\alpha} \left( z_2 \right) T_\delta \left( k , \eta_1 \right) T_v \left( k ,\eta_2 \right)
-  {\mathcal{R}_Q} \left( z_1 \right) b_\alpha \left( z_2 \right) T_v \left( k , \eta_1 \right) T_\delta \left( k ,\eta_2 \right) 
 \right] 
 \nonumber \\
 &&
 + 
 \int \frac{dk}{2 \pi^2} k^3  P_R \left( k \right)
 \left[ - \frac{3}{5} j_1 \left( k x \right)  + \frac{2}{5} j_3 \left( k x \right) \right]
 \left[ 
 \frac{{\mathcal{R}_\alpha} \left( z_2 \right)}{\HH\left( z_1 \right)}  
-  \frac{{\mathcal{R}_Q} \left( z_1 \right) b_v \left( z_2 \right) }{\HH\left( z_2 \right) }
 \right] T_v \left( k , \eta_1 \right) T_v \left( k ,\eta_2 \right)
  \nonumber \\
 &&
 + 
 \int \frac{dk}{2 \pi^2} k^2  P_R \left( k \right)
 \left[ \frac{1}{3} j_0 \left( k x \right)  - \frac{2}{3} j_2 \left( k x \right) \right] 
{\mathcal{R}_Q} \left( z_1 \right) 
{\mathcal{R}_\alpha} \left( z_2 \right)  
 T_v \left( k , \eta_1 \right) T_v \left( k ,\eta_2 \right)
\eea
with
\bea
\label{R_Q}
\mathcal{R}_Q &=& \frac{{\dot\HH}}{\HH^2}+\frac{2-5s}{r_S\HH} +5s-f^Q_{\rm evo} \, ,
\\
\label{R_lya}
\mathcal{R}_\alpha &=&  b_R \left( 3  + \frac{\dot \HH}{\HH^2} - f_\text{evo}^{\elya}\right) \, ,
\eea
where we use a biased expression for ${\mathcal R}_\alpha$.
And the potential term is
\bea
 \xi_{Q \alpha}^{\text{potential}} & \equiv&
 \langle \left[ 
 b_Q \delta+ \HH^{-1} \dd_r(\ndv) + {\mathcal{R}_Q}  \ndv + {\mathcal{P}_Q}
\right] \left( \bn, z_1 \right) 
 \left[ b_\alpha \delta+ b_v \HH^{-1} \dd_r(\ndv)  + {\mathcal{R}_\alpha} \ndv + {\mathcal{P}_\alpha} \right]\left( \bn , z_2 \right) 
\rangle 
\nonumber 
\\
&&
-  \xi_{Q \alpha}^\text{newt}-  \xi_{Q \alpha}^{ \text{Doppler}}
\nonumber \\
&=&
\int \frac{dk}{2 \pi^2} k^2 P_R \left( k \right) j_0 \left( k x \right)
 \left[ b_Q\left( z_1 \right)  T_\delta \left( k , \eta_1 \right) T_{{\mathcal{P}_\alpha}} \left( k , \eta_2 \right) +
b_\alpha \left( z_2 \right)  T_{{\mathcal{P}_Q}} \left( k , \eta_1 \right)  T_\delta \left( k , \eta_2 \right) 
 \right]
 \nonumber 
 \\
 &&+
 \int \frac{dk}{2 \pi^2} k^3 P_R \left( k \right)\left[ - \frac{1}{3} j_0 \left( k x \right) + \frac{2}{3} j_2 \left( k x \right) \right]
 \nonumber \\
 && \qquad 
 \left[ \frac{1}{\HH \left( z_1 \right) } T_v \left( k , \eta_1 \right) T_{{\mathcal{P}_\alpha}} \left( k , \eta_2 \right) +
 \frac{b_v \left( z_2 \right) }{\HH \left( z_2 \right) }
 T_{{\mathcal{P}_Q}} \left( k , \eta_1 \right)  T_v \left( k , \eta_2 \right) 
 \right] 
\nonumber 
 \\
 &&+
 \int \frac{dk}{2 \pi^2} k^2 P_R \left( k \right)  j_1 \left( k x \right)  
 \left[ {\mathcal{R}_Q} \left( z_1 \right) T_v \left( k , \eta_1 \right) T_{{\mathcal{P}_\alpha}} \left( k , \eta_2 \right) -
{\mathcal{R}_\alpha} \left( z_2 \right) 
 T_{{\mathcal{P}_Q}} \left( k , \eta_1 \right)  T_v \left( k , \eta_2 \right) 
 \right] 
 \nonumber 
 \\
 &&+
  \int \frac{dk}{2 \pi^2} k^2 P_R \left( k \right) j _0 \left( k x \right) T_{{\mathcal{P}_Q}} \left( k , \eta_1 \right)  T_{{\mathcal{P}_\alpha}} \left( k , \eta_2 \right)
\eea
where we have used the following definitions
\bea \label{PQ}
{\mathcal{P}_Q} &=& \left( 5s -2\right) \Phi+ \left(  1 +\frac{{\dot\HH}}{\HH^2}+\frac{2-5s}{r_S\HH} +5s-f^Q_{\rm evo} \right) \Psi + \HH^{-1} \dot \Phi  + \left( f^Q_\text{evo} - 3 \right) \HH v \, , \qquad \\
{\mathcal{P}_\alpha} &=& b_R  \left[ \left( 3  + \frac{\dot \HH}{\HH^2} -  f_{\rm evo}^{\elya} \right) \Psi+  \HH^{-1} \dot \Phi + \left( f_{\rm evo}^{\elya}- 3  \right) \HH v \right] 
\label{Palpha} \, ,
\eea
and $T_{{\mathcal{P}_Q}} $ and $T_{{\mathcal{P}_\alpha}}$ denote the respective transfer functions normalized with the primordial curvature perturbation.
The last term in Eq.~(\ref{xi_rel}) denotes the contribution of the
non-local terms, and their cross-correlation with all the other
terms. In particular this includes the correlation between QSO
magnification with all (local and non-local) \lya\ relativistic
effects and the correlation between relativistic non-local terms
(e.g. ISW, Shapiro time delay) with
everything else. Again, as we did for the cosmic magnification we
prefer to use
Eq.~(\ref{2point_cl}) to compute them. 

 \begin{figure}[!ht]
  \centering
  \includegraphics[width=1.0\linewidth]{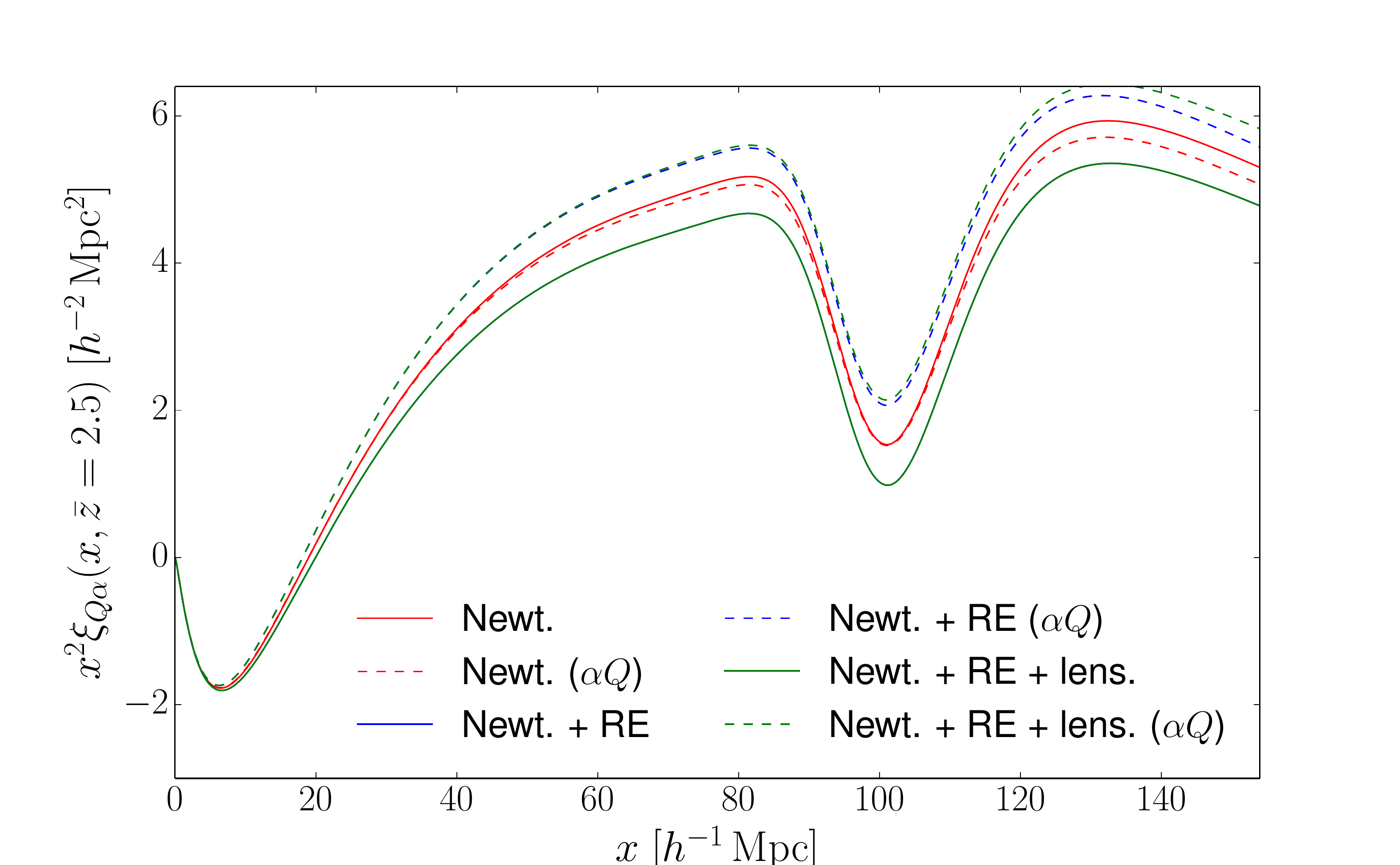}
  \caption{
    The figure shows the Quasar-\lya\ cross-correlation
    function, along the line of sight. Different colours correspond to different contributions:
    red - Newtonian terms (Newt.), blue - Newtonian and relativistic
    effects (RE), green -
    Newtonian, relativistic and lensing terms. The full line and dashed
    line correspond to the cross-correlations $\xi_{Q \alpha}$ and
    $\xi_{\alpha Q}$ respectively. The asymmetry due to the bias
    evolution, relativistic effects and lensing produce a split of the
    cross-correlation function around the symmetric part. The dominant
    effect due to the peculiar velocities of the tracers (Doppler
    effect) is large on all scales above
    $20\;\mathrm{h^{-1}\,Mpc}$ while the lensing and redshift
    evolution effects grow significantly towards larger scales.
  }
  \label{fig:xi_aQ_full}
\end{figure}

Fig.~\ref{fig:xi_aQ_full} shows different contributions to the overall
cross-correlation function. The red line shows the contribution from
the Newtonian terms ($\xi_{\alpha Q}^{\text{Newt}}$), which are the
most standard ones considered in the large-scale correlation
analysis. The largest correction to the signal comes from the
relativistic effects ($\xi_{\alpha Q}^{\text{RE}}$), more specifically
from the Doppler terms (see the subsection on anti-symmetric part for
details). The signal from the relativistic effects is nearly constant for
all scales above $\sim 40\;h^{-1}\mathrm{Mpc}$, and is of around
$10\%$ compared to the pure Newtonian terms. Further we show that the
lensing contribution only gives a small correction at large
scales. Dashed lines correspond to the cross-correlation
$\xi_{\alpha Q}$ , where the positions of \lya\ and QSO have been
exchanged (see Fig.~\ref{cartoon}). This exchange is equivalent to the
transformation $x \rightarrow -x$.

The correction due to relativistic effects is much larger than anticipated from the
galaxy surveys, which is due to two effects: firstly, the difference in biases
is much larger for QSO and \lya\ tracers than for two galaxy
populations, and secondly, the overall signal from the Newtonian terms
is larger as well. The calculations for two galaxy populations have
already been shown in ~\cite{Bonvin:2013ogt}.

\subsection{Local term expansion
in $\HH/k$}
\label{sec:exp}

In our derivation we have been as general as possible and our results
are valid for any metric theory of gravity provided that photons move
along geodesic and photon number is conserved. Since we describe geometrically our observable, the theory of gravity can modify the transfer functions only\footnote{We remark that expressions for bias factors implicitly assume a theory of gravity.}, as long as the previous assumptions are valid. We have also assumed that sources move along geodesics, but this assumption can be easily relieved. 
In order to give the flavor of the amplitude in different regimes of
the effects previously computed in a relativistic framework, we
replace the transfer functions with their counterparts in a
'newtonian' framework. This oversimplification will give us the
physical understanding of the effects shown in the next
sections. Therefore, we apply the following substitutions 
\bea \label{counting_scheme_start}
T_v \left( k , \eta \right) &\rightarrow& - \frac{\HH}{k} f T_\delta \left( k , \eta \right)  \\
T_\Psi \left( k , \eta \right) &\rightarrow& - \frac{3}{2} \frac{\HH^2}{k^2} \Omega_M T_\delta \left( k , \eta \right)  \\
T_\Phi \left( k , \eta \right) &\rightarrow& - \frac{3}{2} \frac{\HH^2}{k^2} \Omega_M T_\delta \left( k , \eta \right)\\
T_{\dot \Phi} \left( k , \eta \right) &\rightarrow& - \frac{3}{2} \frac{\HH^3}{k^2} \Omega_M \left( f - 1 \right) T_\delta \left( k , \eta \right)
\label{counting_scheme_end}
\eea
where the growth factor can be approximated in perturbation theory in $\Lambda$CDM by $f(z) = \Omega_M (z)^{0.6}$.

By evaluating all the cosmological parameters and biases at the mean redshift $z_\text{mean}$ we find
\bea
\label{newt_simp}
\xi_{Q \alpha}^\text{newt} &\sim&  b_Q  b_\alpha \int \frac{d k }{2 \pi^2} k^2  P \left( k \right) j_0 \left(   k x\right)   
\nonumber \\
&&+ b_v   \int \frac{dk }{2 \pi^2} k^2 f^2 P \left( k \right)
 \left[ \frac{1}{5} j_0  \left(   k x \right)  - \frac{4}{7} j_2 \left(   k x \right) + \frac{8}{35} j_4 \left(   k x \right)\right]
 \nonumber \\
 && + \left( b_Q b_v + b_\alpha \right) 
 \int \frac{dk}{2 \pi^2} k^2  f P \left( k \right) 
 \left[ \frac{1}{3} j_0 \left(   k x \right)  - \frac{2}{3} j_2 \left(   k x \right)  \right]
, \\
\label{doppler_simp}
\xi_{Q \alpha}^\text{Doppler} &\sim&  \left( - b_Q  {\mathcal{R}_\alpha} + {\mathcal{R}_Q} b_\alpha \right)\int \frac{dk}{2 \pi^2} k^2  f P \left( k \right)
 j_1 \left( k x \right) \frac{\HH}{k}
 \nonumber \\
 &&
 + 
\left( - {\mathcal{R}_\alpha} + {\mathcal{R}_Q} b_v \right)  \int \frac{dk}{2 \pi^2} k^2 f^2 P \left( k \right)
 \left[  \frac{3}{5} j_1 \left( k x \right)  - \frac{2}{5} j_3 \left( k x \right) \right]
\frac{\HH}{k}
  \nonumber \\
 &&
 + 
 {\mathcal{R}_\alpha} {\mathcal{R}_Q}
 \int \frac{dk}{2 \pi^2} k^2  f^2 P \left( k \right)
 \left[ \frac{1}{3} j_0 \left( k x \right)  - \frac{2}{3} j_2 \left( k x \right) \right] 
\left( \frac{\HH}{k} \right)^2 \, , \\
\label{pot_simp}
\xi_{Q \alpha}^\text{potentials} &\sim& \left( b_Q { \tilde{\mathcal{P}}_\alpha} + { \tilde{\mathcal{P}}_Q} b_\alpha \right)
\int \frac{dk}{2 \pi^2} k^2 P \left( k \right) j_0 \left( k x \right) \left( \frac{\HH}{k}\right)^2
 \nonumber 
 \\
 &&+
 \left(  { \tilde{\mathcal{P}}_\alpha} + { \tilde{\mathcal{P}}_Q} b_v \right)
 \int \frac{dk}{2 \pi^2} k^2 f P   \left( k \right)\left[  \frac{1}{3} j_0 \left( k x \right) - \frac{2}{3} j_2 \left( k x \right) \right]
\left( \frac{\HH}{k}\right)^2
\nonumber 
 \\
 &&+
 \left( - {\mathcal{R}_Q} { \tilde{\mathcal{P}}_\alpha} + { \tilde{\mathcal{P}}_Q} {\mathcal{R}_\alpha} \right)
 \int \frac{dk}{2 \pi^2} k^2 f P \left( k \right)  j_1 \left( k x \right)  \left( \frac{\HH}{k}\right)^3
  \nonumber 
 \\
 &&+
 { \tilde{\mathcal{P}}_\alpha} { \tilde{\mathcal{P}}_Q}
  \int \frac{dk}{2 \pi^2} k^2  P \left( k \right) j _0 \left( k x \right)\left( \frac{\HH}{k}\right)^4
 \eea
where $P(k) = T_\delta \left( k , \eta \right)^2 P_R (k)$ denotes the matter power spectrum at the mean redshift. ${ \tilde{\mathcal{P}}_Q}$ and ${ \tilde{\mathcal{P}}_\alpha}$ are the pre-factors defined through the transformations~(\ref{counting_scheme_start} - \ref{counting_scheme_end}), namely
\bea
T_{\mathcal{P}_Q}\left( k ,\eta \right) &\rightarrow &{ \tilde{\mathcal{P}}_Q} \frac{\HH^2}{k^2}T_\delta \left( k ,\eta \right) \, ,\\
T_{\mathcal{P}_\alpha}\left( k ,\eta \right) &\rightarrow& { \tilde{\mathcal{P}}_\alpha} \frac{\HH^2}{k^2} T_\delta \left( k ,\eta \right) \, ,
\eea
and they can be explicitly computed
\bea
\tilde {\mathcal{P}}_Q &=&\frac{3}{2} \Omega_M \left( 2 - f + f^Q_\text{evo} - \frac{\dot \HH}{\HH^2}+ \frac{2 f}{3 \Omega_M} \left( 3 - f^Q_\text{evo}\right) - 10 s + \frac{5s -2}{\HH r} \right) \,  ,  \\
\tilde {\mathcal{P}}_\alpha &=& b_R \frac{3}{2} \Omega_M
\left( -2 + f_{\rm evo}^{\elya} - \frac{\dot \HH}{\HH^2} - f + \frac{2 f}{3 \Omega_M} \left( 3 - f_\text{evo}^{\elya} \right)\right) \, .
\eea
We remark that the transformation~(\ref{counting_scheme_start} - \ref{counting_scheme_end}) has been performed by using Newtonian equations. Nevertheless, it is well known that, for CDM, linearized Einstein equations agree with Newtonian dynamics if perturbations are considered in specific gauges, namely the density perturbation in synchronous gauge and the velocity in newtonian gauge. Since we have chosen to express perturbations in this combination of gauges, the expressions here derived include correctly the GR dynamics, as long as the approximation that CDM is the only species which contributes to perturbations is valid.

According to this scheme, we note that all the newtonian terms
contribute at the same parametrical order. The Doppler term,
Eq.~(\ref{doppler_simp}), contains the leading correction of order
$\HH/k$ to the newtonian approximation. This term is non-vanishing
only if we use two different tracers or probes. With a single tracer 
the leading correction would be of the order $\left(\HH/k\right)^2$, 
and the amplitude of these terms would be much more suppressed. 
The leading corrections have also a different parity under the 
exchange $x \leftrightarrow -x$. Because of that, they contribute 
only to the antisymmetric part of the correlation function, while 
newtonian terms (with biases and cosmology evaluated at the mean
redshift $z_\text{mean}$) 
lead to a symmetric correlation function only. 
This argument does not only show the importance of using different 
probes to try to measure relativistic correction, as already shown 
in \cite{McDonald:2009ud,Yoo:2012se,croft13,Bonvin:2013ogt}, but it also 
indicates that when we consider cross-correlations between different 
probes we need to carefully consider effects which might be 
neglected with a single tracer, as shown 
in~\cite{Yoo:2012se,Alonso:2015uua}. For simplicity, in this 
counting scheme we did not include integrated effects like cosmic 
magnification, ISW and Shapiro time delay effects. It is known that 
cosmic magnification can be of the same parametrical order of density 
and redshift space distortion perturbations, 
see e.g.~\cite{DiDio:2013sea, Raccanelli:2013gja, Montanari:2015rga}
and we do include it in the full relativistic analysis in next
sections. In the full analysis there is also a contribution to the 
antisymmetric correlation function due to newtonian terms because 
of the biases and cosmological parameter evolutions, that have been 
neglected in this simple approach.

\subsection{Symmetric correlation function}
\label{sec:sym}

The correlation function~(\ref{2point}) can decompose in a symmetric and anti-symmetric part. In this section we are interested in the symmetric correlation function
\be
\xi_{(Q\alpha)} \left(z_1, z_2, \theta \right) = \frac{\xi_{Q\alpha} \left(z_1, z_2, \theta \right)  + \xi_{\alpha Q} \left(z_1, z_2, \theta \right) }{2} 
\ee
Up to the redshift dependence of transfer functions or bias factors, only the correlation functions which involve even spherical Bessel functions contribute to the symmetric part of the correlation function.

 \begin{figure}[!h]
  \centering
  \includegraphics[width=1.0\linewidth]{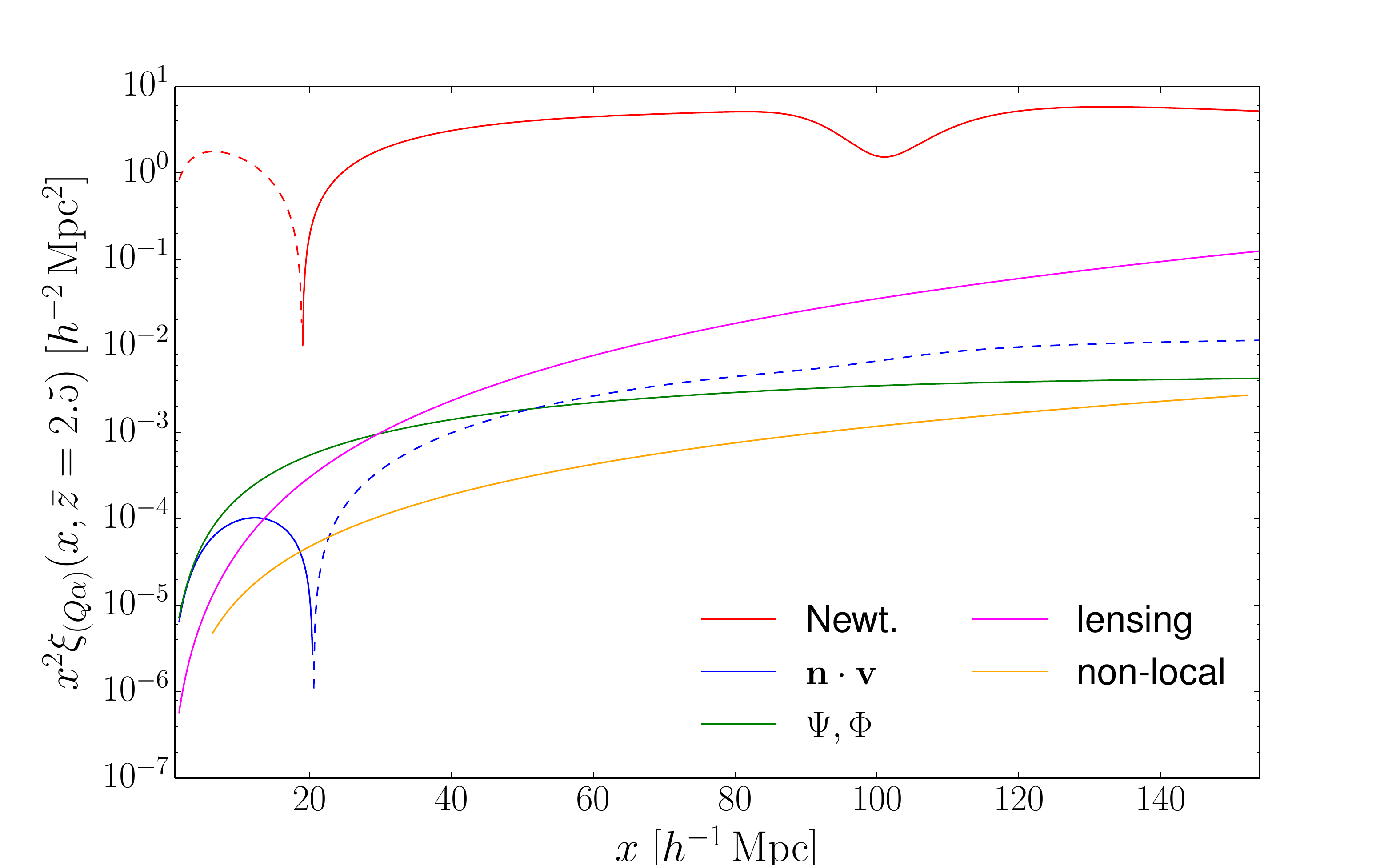}
  \caption{
    The figure shows the symmetric part of the Quasar-\lya\ cross-correlation
    function, along the line of sight. Different colours correspond to different contributions:
    red - Newtonian terms (Newt.), blue - Doppler terms, green -
    potential terms, magenta - lensing terms and orange - non-local
    terms. The full line and dashed
    line correspond to the positive and negative values of the
    cross-correlations respectively. Newtonian terms dominate the
    signal in the symmetric part of the cross-correlation
    function. Note that non-local terms were not computed at very
    small scales due to numerical issues, however they are very
    small. The lensing contribution (magenta) is increasing towards
    large scales which has been subject to many discussions in the
    literature (see text for details).
  }
\label{fig:sym}
\end{figure}

Figure~\ref{fig:sym} shows the symmetric contribution to the
QSO-\lya\ cross-correlation function. As expected the Newtonian terms
(density and RSD) dominate the symmetric signal, while the
Relativistic terms (Doppler and potential) only slightly contaminate
the result. Similar effect could be observed when one considers only
auto-correlations of one tracer. It is interesting to note that while
both the Doppler and potential terms in the symmetric form are of the
same order of magnitude in the expansion of $k$ ($\sim(\HH/k)^2$), the
values of the cosmological and bias prefactors are different - mainly the
large contribution to the Doppler term comes from the large values of
the QSO ($f_{\mathrm{evo}}^Q$) and \lya\ evolution bias
($f_{\mathrm{evo}}^{\elya}$). 
Also, while Doppler terms
are in general larger than the potential terms, the later become the
leading sub-dominant signal at small scales.
At very large scales, on the other hand, the lensing terms of the QSO
start to become the dominant contaminant. This is the main idea behind
the recent papers claiming that using a large window function in the
radial direction will boost the lensing signal~\cite{DiDio:2013sea}. In fact, it is more
about decreasing the overall signal of the density perturbation, which
remain more or less constant in radial direction, compared to the
lensing term which is always increasing with the distance.

The non-local terms are always small compared to the other effects in
the symmetric part of the cross-correlation function. As for the
lensing part of the non-local terms, the relative contribution of it
grows for large (redshift) separations.

\subsection{Anti-symmetric correlation function}
\label{sec:asym}

Analogously to the symmetric part, we can define the anti-symmetric
part of the correlation function

\be
\xi_{[Q\alpha]} \left(z_1, z_2, \theta \right) = \frac{\xi_{Q\alpha}
  \left(z_1, z_2, \theta \right)  - \xi_{\alpha Q} \left(z_1, z_2,
  \theta \right) }{2} \, .
\ee

A non vanishing anti-symmetric correlation function will indicate a violation of the exchange symmetry of the
tracers ($z_1 \leftrightarrow z_2$). Thus,
the interchange of the tracers at the same position would produce
different results, and which tracer is in the background and which in
the foreground would matter. However, this is only true when one is
dealing with differently biased tracers. If, for instance, one
computes the auto-correlation function of one single tracer, the
correlation function will be by definition symmetric. On the other hand two
different effects can break the symmetry when dealing with different
tracers. 
The asymmetry can come either from different value and redshift
evolution of the bias factors or from the intrinsic evolution of the
transfer functions. In the case
of the galaxy surveys that is the case, and several sources of the
asymmetry have been identified and suggested in the literature
(e.g. gravitational redshift, Doppler effects,
etc.)~\cite{McDonald:2009ud,croft13,Bonvin:2013ogt}. 
These effects reflect the fact that there is not a translation
symmetry along the redshift direction, and the asymmetry is generated
along the radial direction. Whereas in the angular direction, if statistical
isotropy and (angular) homogeneity are not broken, any asymmetry can not
be sourced. 

\begin{figure}[!h]
  \centering
  \includegraphics[width=1.0\linewidth]{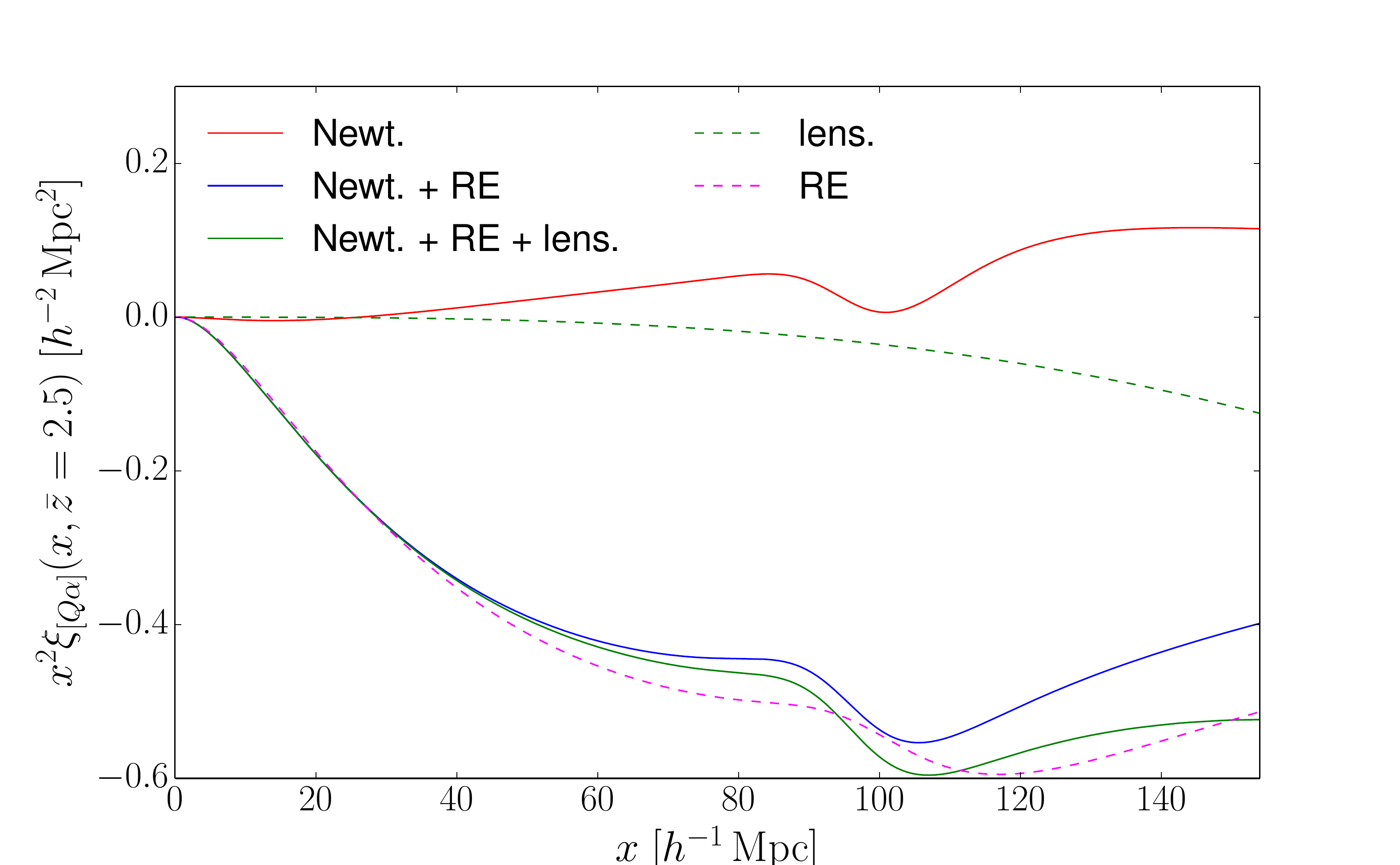}
  \caption{
    The figure shows different contributions to the anti-symmetric
    part of the cross-correlation function, along the line of sight. Full red line shows the
    contribution due to the often neglected redshift evolution of the
    bias factors, dashed magenta line shows the contribution due to
    the relativistic effects and dashed green line contribution of the
    lensing terms. The full blue and green lines represent summing up
    various different terms - redshift evolution and Doppler terms
    (full blue line), and all the effects (full green line). The main
    contributor to the signal are the relativistic effects (mostly
    Doppler terms), while the bias redshift evolution and lensing form
    the sub-dominant corrections of order of $10\%$ of the signal in
    the anti-symmetric cross-correlations.
  }
\label{fig_asym}
\end{figure}

Our calculations in the previous sections show that the main source of
asymmetry of the correlation functions comes from the Doppler terms in
the relativistic expansion (see Fig.~\ref{fig_asym}). 
In our simple expansion, see Eqs.~(\ref{newt_simp},
\ref{doppler_simp}, \ref{pot_simp}), the Doppler term,
Eq.~(\ref{doppler_simp}) contains terms with odd spherical Bessel
functions multiplied by only one factor $\HH/k$. In the newtonian
term, Eq.~(\ref{newt_simp}), all the terms are multiplied by even
Bessel function, and the asymmetry is generated by the redshift
dependence only, while in potential term, Eq.~(\ref{pot_simp}), the
terms with odd Bessel function are suppressed by $\left( \HH/k
\right)^3$.
The largest single contribution is the Doppler -
density correlation which is boosted by the difference in bias factors
(see Eq.~\ref{doppler_simp}). This result is known in the literature
and has been commented on at various times, mostly in the context of
two galaxy populations~\cite{McDonald:2009ud,Bonvin:2013ogt}. In this paper we have
focused on two tracers whose bias factors are as different as
possible: quasar bias being very large, and \lya\ forest flux bias
being even negative. Thus ensuring that the signal from the Doppler
effect is boosted as much as possible.

The first of the sub-dominant signals is the redshift evolution of the
bias factors (see Fig.~\ref{fig_asym}). Usually one approximates the redshift dependent
quantities within a redshift bin with a constant value at the mean
redshift. However, redshift evolution of the bias factors gives a
fairly large contribution to the asymmetry. This contribution mainly
comes from the standard Kaiser terms - indeed one can easily replicate
the results by using well known Eq.~(\ref{newt_simp}) and requiring that
the bias factors for quasar and \lya\ are evaluated not at the mean
redshift $z_{\text{mean}}$ but at $z_{1}$ and $z_{2}$ respectively.
Strictly speaking, this effect should not be considered for biases
only, but for any redshift dependent prefactors (e.g.~cosmological
prefactors of the relativistic effects, growth factors, etc.), however
in our calculations the effects of growth of structure and evolution
of the cosmological parameters (most notably $H(z)$) are very small
compared to the other effects, and are well below the $1\%$ of the
final result. Thus one can safely use the approximation of constant
cosmological parameters evaluated at the mean redshift.

The signal of the bias redshift evolution gives an important
contribution, that needs to be modeled if one wants to measure the
effects to below $10\%$ accuracy. While it can be regarded as a
nuisance signal to the relativistic effects, it requires a precise
knowledge of the bias redshift dependence of both tracers. However,
this information can be acquired from independent and complementary
results, such as auto-correlation functions of each of the tracers,
and can be well constrained.

\begin{figure}[!h]
  \centering
  \includegraphics[width=1.0\linewidth]{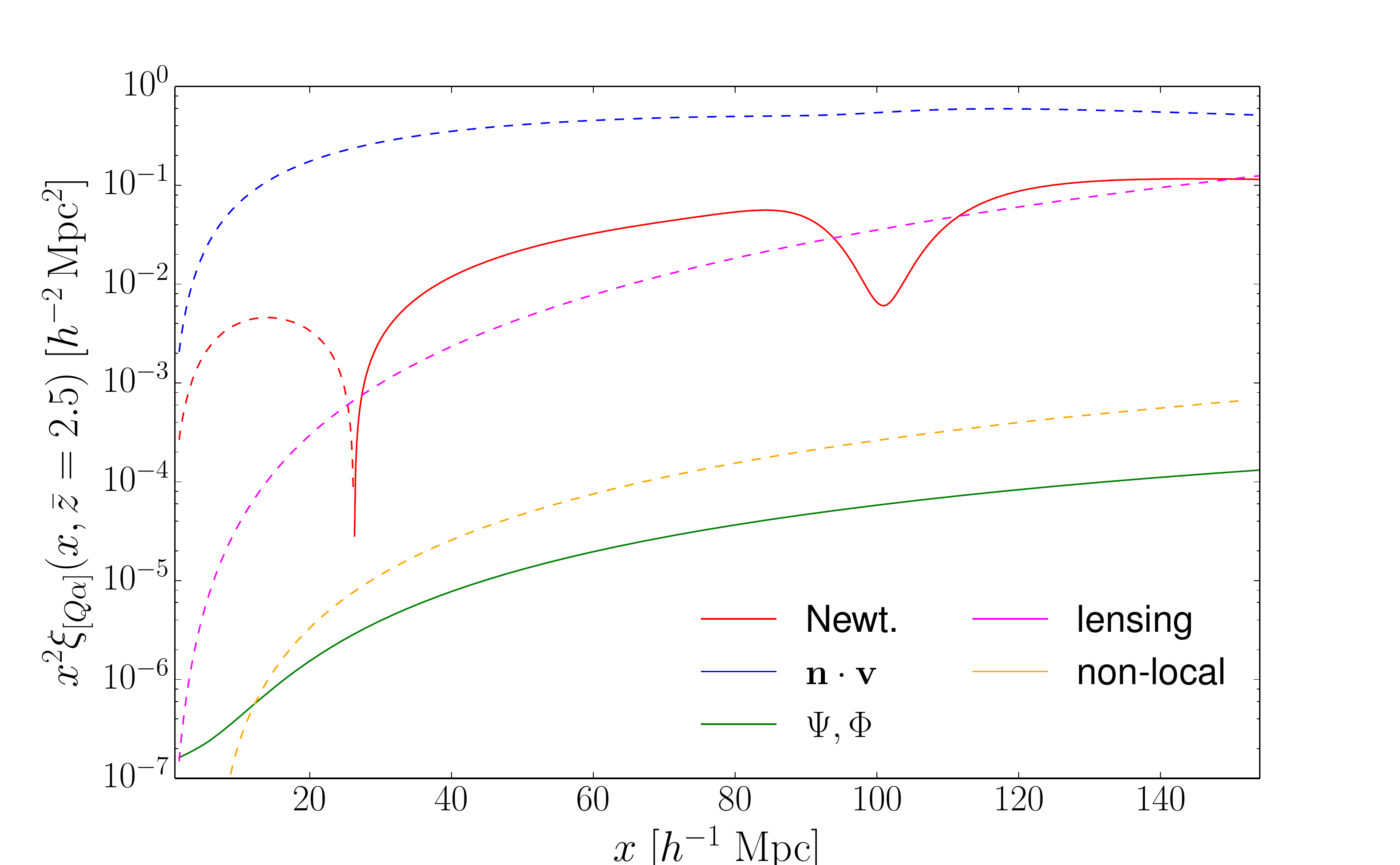}
  \caption{
    The figure shows different contributions to the relativistic
    effects in anti-symmetric part of the cross-correlation
    function, along the line of sight. The blue line shows contributions due to the Doppler
    effects, red line due to potential terms, and green line due to
    all non-local terms. The difference between full and dashed line
    is whether the contribution to $\xi_{[Q\alpha]}$ is positive or
    negative respectively. Compared to the symmetric part the
    potential and Newtonian terms are strongly suppressed.
  }
\label{fig_asym_nonlocal}
\end{figure}

Fig.~\ref{fig_asym_nonlocal} shows the contributions of the
relativistic terms to the anti-symmetric part of the $\xi_{Q\alpha}$,
broken down into its components. It is clear that at all scales the
cross-correlations due to Doppler terms dominate the relativistic
signal, while the potential and non-local parts are small. The next to
leading order corrections in the anti-symmetric part come from the
redshift evolution of the bias factors in the Newtonian terms, an
effect often considered to be too small to be of any importance. In
the case of QSO-\lya\ cross-correlation it is an important
sub-dominant contribution to the anti-symmetric part. Compared to the
symmetric part of the $\xi_{Q \alpha}$ (see Fig.~\ref{fig:sym}) the
potential terms are also suppressed at all scales, even compared to
the non-local terms.

A special note deserves the fact that all the contaminating effects
are small at the BAO scale. This result does not carry any deeper
significance, and is a result of the values of bias factors for the
tracers used in this study. Nevertheless, for this particular case it does allow for an easier
estimation of the relativistic effects at the BAO scale - which is
already a target of many surveys and being a robust feature, very
accurately measured. 

\begin{figure}[!h]
  \centering
  \includegraphics[width=1.0\linewidth]{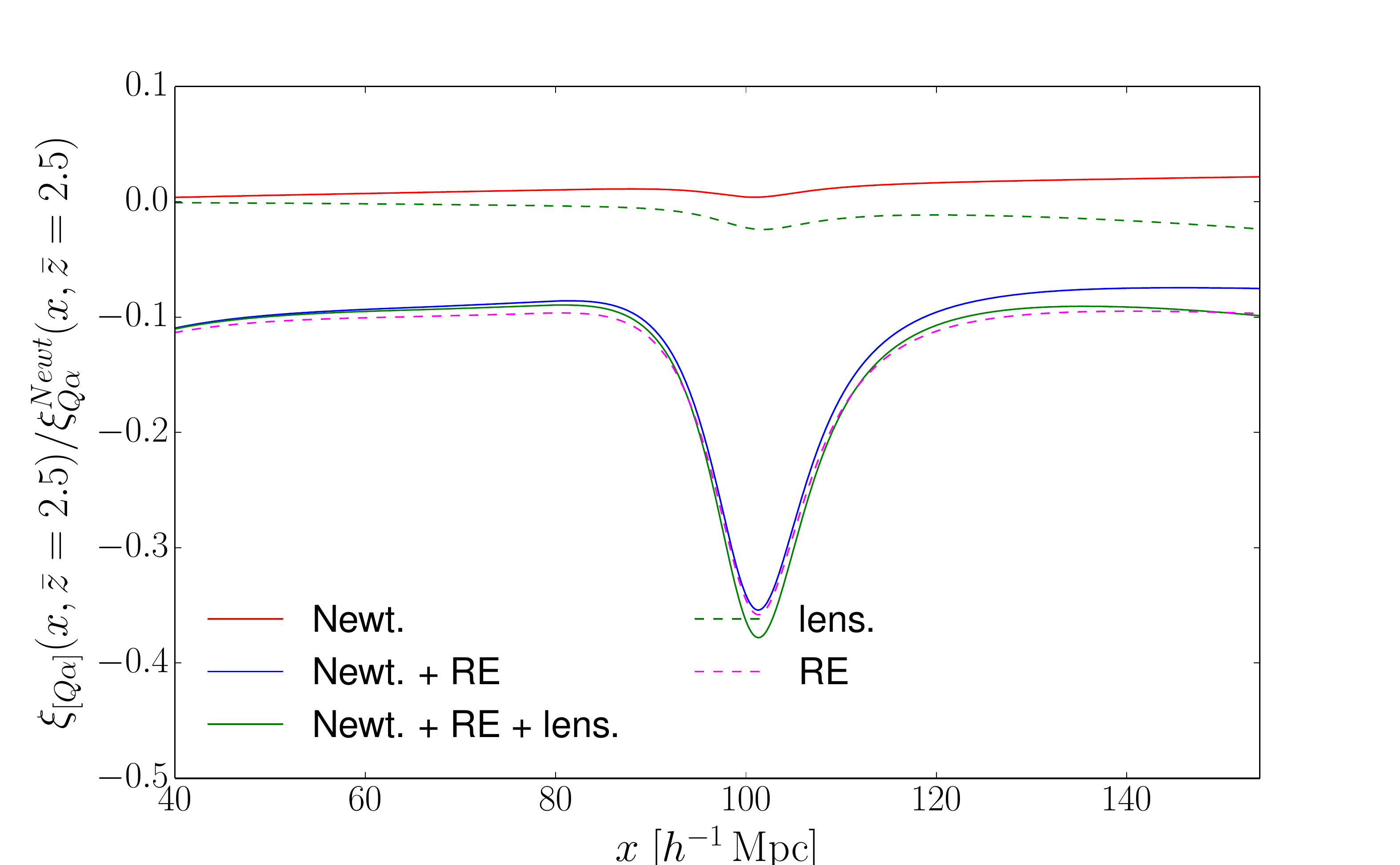}
  \caption{
    The figure shows the relative size of the anti-symmetric
    part of the cross-correlation function, along the line of sight, for different terms. Full red line shows the
    contribution due to the redshift evolution of the
    bias factors, dashed magenta line shows the contribution due to
    the relativistic effects and dashed green line contribution of the
    lensing terms. The full blue and green lines represent summing up
    various different terms - redshift evolution and Doppler terms
    (full blue line), and all the effects (full green line). The
    relativistic effects can be as large as 10\% at all scales above
    $~\sim 40\;h^{-1}\mathrm{Mpc}$,
    compared to the pure Newtonian calculation (including the bias
    redshift evolution).
  }
\label{fig_asym_perc}
\end{figure}

Figure~\ref{fig_asym_perc} shows that the signal
of the relativistic effects is largest at the BAO scale, compared to
the cross-correlation function computed just with Newtonian terms. The size
of the asymmetry is around 10\%. One should note that even though the
ratio reach around 30\% at the BAO scale this is an artificial
enhancement of the signal since the cross-correlation function drops
nearly to zero. The relativistic effects are nearly constant on the
scale larger than $\sim 40\;h^{-1}\mathrm{Mpc}$ and contribute to
around 10\% distortion on all scales larger than that.

The second contaminating effect is that of the weak lensing signal of quasars. In
our specific case the lensing signal can be as large as the effects
due to the bias redshift evolution. Similarly to
the bias redshift evolution signal, the lensing signal becomes more
important on larger scales, and requires careful modeling if one is to
measure the overall asymmetry effects to a high precision.



\begin{figure}[!h]
  \centering
  \includegraphics[width=1.0\linewidth]{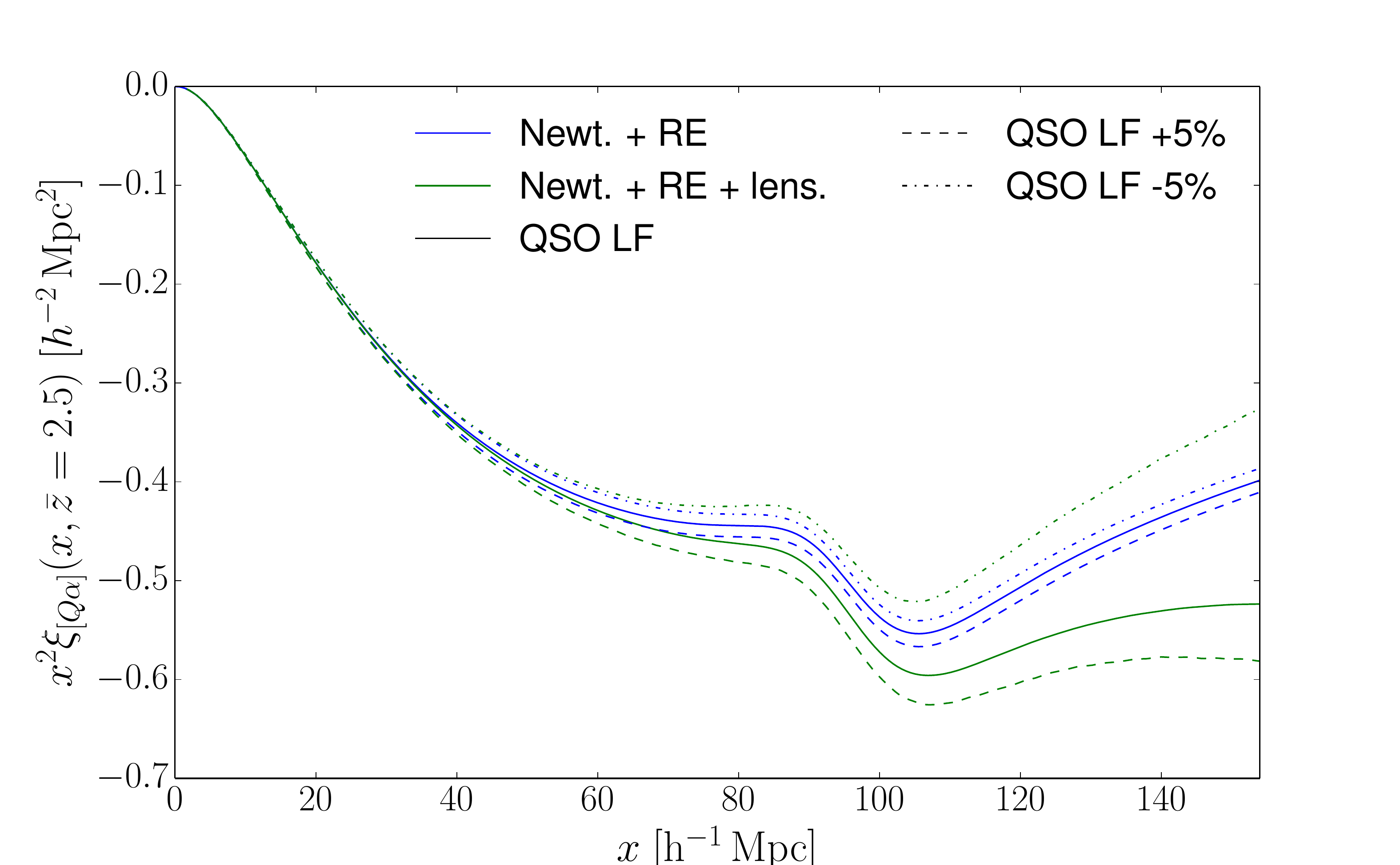}
  \caption{
    The asymmetry part of the QSO-\lya\ cross-correlation
    function, along the line of sight. Blue colour represents Newtonian and relativistic terms,
    while green colour adds lensing to them. The different line shapes
    correspond to different values of the bias parameters: full line
    shows the mean values used in our calculations, dashed line shows
    the values ($s = 0.256883$, $f_{\text{evo}}^Q = 6.19163$) for $+5\%$
    change to the QSO LF parameters and dot-dashed the values ($s =
    0.450115$, $f_{\text{evo}}^Q = 4.74261$) for $-5\%$ change to the
    QSO LF parameters. Even at the BAO scales - the values of the
    evolution and magnification bias can affect the result by $10\%$
    and need to be modeled carefully.
  }
  \label{fig:asym_bias}
\end{figure}

However, the weak lensing effect is proportional to the magnification
bias (and on density bias, since the leading term is $<\kappa
\delta>$), and thus its signal is strongly dependent on this nuisance
parameter. We caution that magnification bias needs to be well known,
and possibly constrained from independent results, to avoid the
contamination of the signal. The imprecise knowledge of the bias
factor affects larger scales the most, it remains important
contaminant even at the BAO scales, where it affects the amplitude of
the BAO peak, but not its position. Figure \ref{fig:asym_bias} shows
the effects on the asymmetry produced if we change the measured
parameters of the Quasar Luminosity Function used to derive the
evolution and magnification bias parameters. While the effect of the
evolution bias parameter is smaller, it is also degenerate with the
effect of the magnification bias at larger scales. However, the change of
parameters by 5\% shown in the plot is conservative, and future
surveys and current surveys will be able to estimate the quasar
luminosity function more accurately and hence provide a better
estimate on the parameter $f_{\rm evo}^Q$.

In a standard (single tracer) large scale observable, relativistic
effects (except cosmic magnification) are naively suppressed on
sub-Hubble scales by the variance induced by the leading newtonian
terms and are limited by cosmic variance on large scales. Because of
these limitations, all forecasts, see
e.g.~\cite{Yoo:2012se,Yoo:2013zga,Alonso:2015uua}, predict that they
are not observable in current and future planned surveys. 
Nevertheless the antisymmetric correlation function, and more generically the
dipole of the 2-point function, represents an independent observable
whose leading signal is determined, as we have shown, by relativistic
corrections. Hence the detection of relativistic effects is not
anymore limited by cosmic variance but it becomes a shot-noise limited
observable. Therefore, relativistic effects can be detected on
scale much smaller than Hubble size. Not only larger scale experiment,
but also advanced data analysis technique like
multi-tracers~\cite{Seljak:2008xr}, shot-noise
canceling~\cite{Seljak:2009af} and generically correlations between
different tracers will probably require to include relativistic
effects to correctly interpret the observation and to avoid to
introduce any theoretical bias.

The last effect that can potentially contaminate the cosmological
signal in the relativistic terms is the uncertainty in the
$f_{\mathrm{evo}}^{\elya}$. Its exact value depends on the assumption
about the redshift evolution of the mean gas temperature and the
ionizing UV background.

Arguably one might worry that such a redshift evolution will change
the value of $f_{\text{evo}}^{\elya}$. While that is in principle true,
the change is not significant. 

\begin{figure}[!h]
  \centering
  \includegraphics[width=1.0\linewidth]{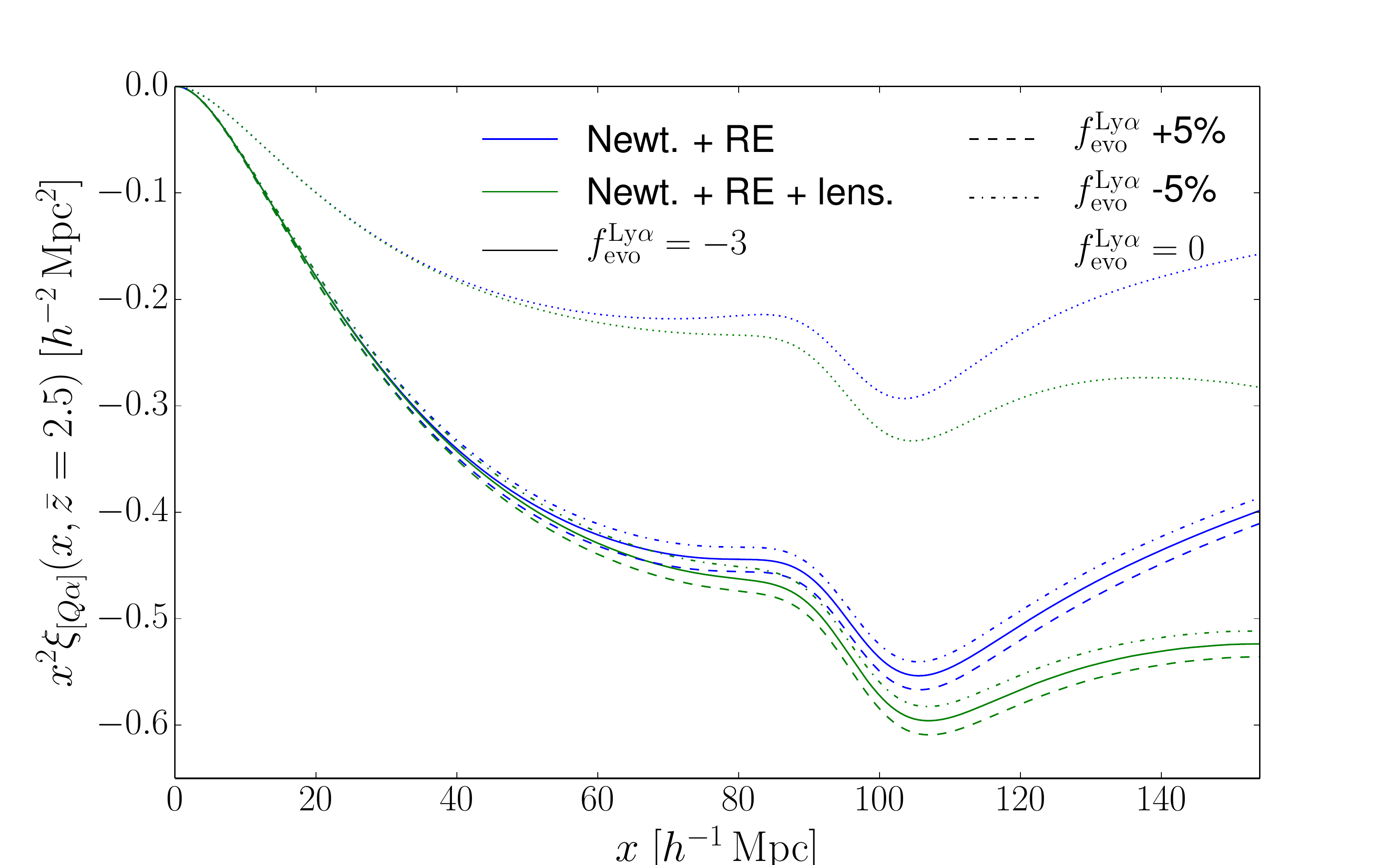}
  \caption{
    The asymmetry part of the QSO-\lya\ cross-correlation
    function along the line of sight. Blue colour represents Newtonian and relativistic terms,
    while green colour adds lensing to them. The different line shapes
    correspond to different values of the \lya\ evolution bias parameters: full line
    shows the mean values used in our calculations, dashed line shows
    the values ($f_{\text{evo}}^{\elya} = -3.15$) for $+5\%$
    change to value adopted in this paper and dot-dashed the values 
    ($f_{\text{evo}}^{\elya} = -2.85$) for $-5\%$ change to the mean
    value. The changes in the $\pm 5\%$ are very small, and seem to be
    in agreement with measured redshift evolutions
    (\cite{bolton04,puchwein14}). The dotted line corresponds to a
    zero evolution bias, which is a very extreme case and would
    require a strong, increasing evolution in the mean temperature
    with redshift, and no evolution of the UV background.
  }
  \label{fig:asym_fevolya}
\end{figure}

The current measurements of the UV background evolution
(\cite{bolton04}) suggest that $\Gamma_{\gamma,HI}$ is a constant in
the redshift range of $z=2-4$. While both the semi-analytical models
and numerical simulations suggest an evolving $\Gamma_{\gamma,HI}$,
the predicted redshift evolution is decreasing with redshift in the
redshift range in question, and is consistent with the measurements. A
decreasing redshift evolution in the UV background would further
decrease $f_{\text{evo}}^{\elya}$, resulting in stronger overall signal
in the asymmetric part of the cross-correlation function (see the
corresponding $+5\%$ line in Fig.~\ref{fig:asym_fevolya}).

On the other hand the measurements of the mean temperature show that
it is increasing with redshift at $z=2-3$ \cite{puchwein14}. 
Thus the evolution of $T_0$ would increase the value of
$f_{\text{evo}}^{\elya}$ and decrease the signal of (primarily) Doppler
term (see the corresponding $-5\%$ line in Fig.~\ref{fig:asym_fevolya}). 

In principle the models and observations suggest a stronger evolution
in the mean temperature which could lower $f_{\text{evo}}^{\elya}$,
but the current estimates (\cite{bolton04,puchwein14}) suggest that
the true value is within $\pm 5\%$ of the value used in this paper
($f_{\mathrm{evo}}^{\elya} = -3$).

\subsection{Signal-to-noise analysis}
\label{sec:SN}

We have performed a simple signal-to-noise analysis to show the
relevance of the relativistic effects in the future surveys. We chose
to follow a prescription in the Fourier space, used in
\cite{McDonald:2009ud}. While we note that our results and calculations
have been done in real space, the Fourier space $S/N$ calculation is
still valid. Indeed, as long as we neglect systematic effects, if we
cannot observe the effect in Fourier space, we cannot observe it in the real-space, and vica versa.

For the purpose of this calculation we have considered only the
largest relativistic correction (Doppler effect) as the source of the
new signal. We have shown that in our case this is a good
approximation, moreover, it is the result of cutting the 
expansion in $\HH/k$ at second order. Thus the power spectra for the
\lya\ forest and QSO simplify to the well known Kaiser formula
\bea
\label{P_auto}
P_\alpha(k,\mu) &=& \left( b_\alpha + b_v \mu^2 f \right)^2 P(k) +
{\cal O}\left(\HH^2/k^2\right) \notag \\
P_Q(k,\mu) &=& \left( b_Q + \mu^2 f \right)^2 P(k) +
{\cal O}\left(\HH^2/k^2\right),
\eea
where the angle follows the notation of this paper ($\mu \equiv -{\bf n}
\cdot {\hat {\bf k}}$).

The cross-power can be obtained by combining
Eqs.~(\ref{delta_Q}) and~(\ref{Lya_Delta}), dropping all terms of order ${
  \HH^2/k^2}$ or higher and performing a Fourier transform. The result
can be written as
\be
\label{P_cross}
P_{\alpha Q}(k,\mu) = \left( b_\alpha + b_v \mu^2 f + i {\mathcal R}_\alpha
\mu \frac{\HH}{k} \right)  \left( b_Q + \mu^2 f - i {\mathcal R}_Q \mu \frac{\HH}{k}
\right) P(k),
\ee
where ${\mathcal R}_Q$ and ${\mathcal R}_\alpha$ are given by
Eq.~(\ref{R_Q}) and~(\ref{R_lya}) respectively. The values at $z=2.5$
for these two terms are ${\mathcal R}_Q(z=2.5) \approx -4.4$ and 
${\mathcal R}_\alpha \approx -1.17$. The largest contribution comes
from the evolution bias for both QSOs and \lya\ forest. However, an
important role is also played by the bias factor of relativistic terms ($b_R$), which
lowers the value of ${\mathcal R}_\alpha$ and makes both terms of the
same sign.

We have shown that the relativistic signals dominantly show in the
anti-symmetric part of the cross-correlation function. Translating
this effect to the Fourier space, it corresponds to the imaginary part
of the cross power spectrum \cite{McDonald:2009ud}. Splitting the
cross power into the relevant contributions we define the real and
imaginary part as
\bea
P_{\alpha Q}^R(k,\mu) &=& \left( b_\alpha + b_v \mu^2 f \right) \left(
b_Q + \mu^2 f \right) P(k) \notag \\
P_{\alpha Q}^I(k,\mu) &=& \frac{\HH}{k} \mu f \left[ {\mathcal R}_\alpha
  \left(b_Q + \mu^2 f\right) - \left(b_\alpha + b_v \mu^2 f\right) {\mathcal R}_Q \right] P(k).
\eea
The variance of the imaginary part, as shown by \cite{McDonald:2009ud}, is 
\be
\sigma^2_I(k,\mu) = \left.P_{\alpha Q}^I \right.^2 + \frac{1}{2}
N_\alpha P_Q + \frac{1}{2} N_Q P_\alpha + \frac{1}{2} N_\alpha N_Q,
\ee
where $N_\alpha$ and $N_Q$ are the noise power spectra for
\lya\ forest and QSOs respectively.
The signal-to-noise estimator for a survey of a volume $V$ is an
integral over the modes of the per mode $\left(S/N\right)_k$ ratio:
\be
\left(\frac{S}{N}\right)^2 = \frac{V}{8 \pi^2}
\int_{k_{\text{min}}}^{k_{\text{max}}} dk k^2 \int_{-1}^1 d\mu
\left(\frac{P_{\alpha Q}^I(k,\mu)}{\sigma_I(k,\mu)}\right)^2,
\ee
where the large scale mode $k_{\text{min}}$ is determined by the
volume of the survey, i.e. $k_{\text{min}} \sim 2\pi/V^{1/3}$.

\begin{figure}[!h]
  \centering
  \includegraphics[width=1.0\linewidth]{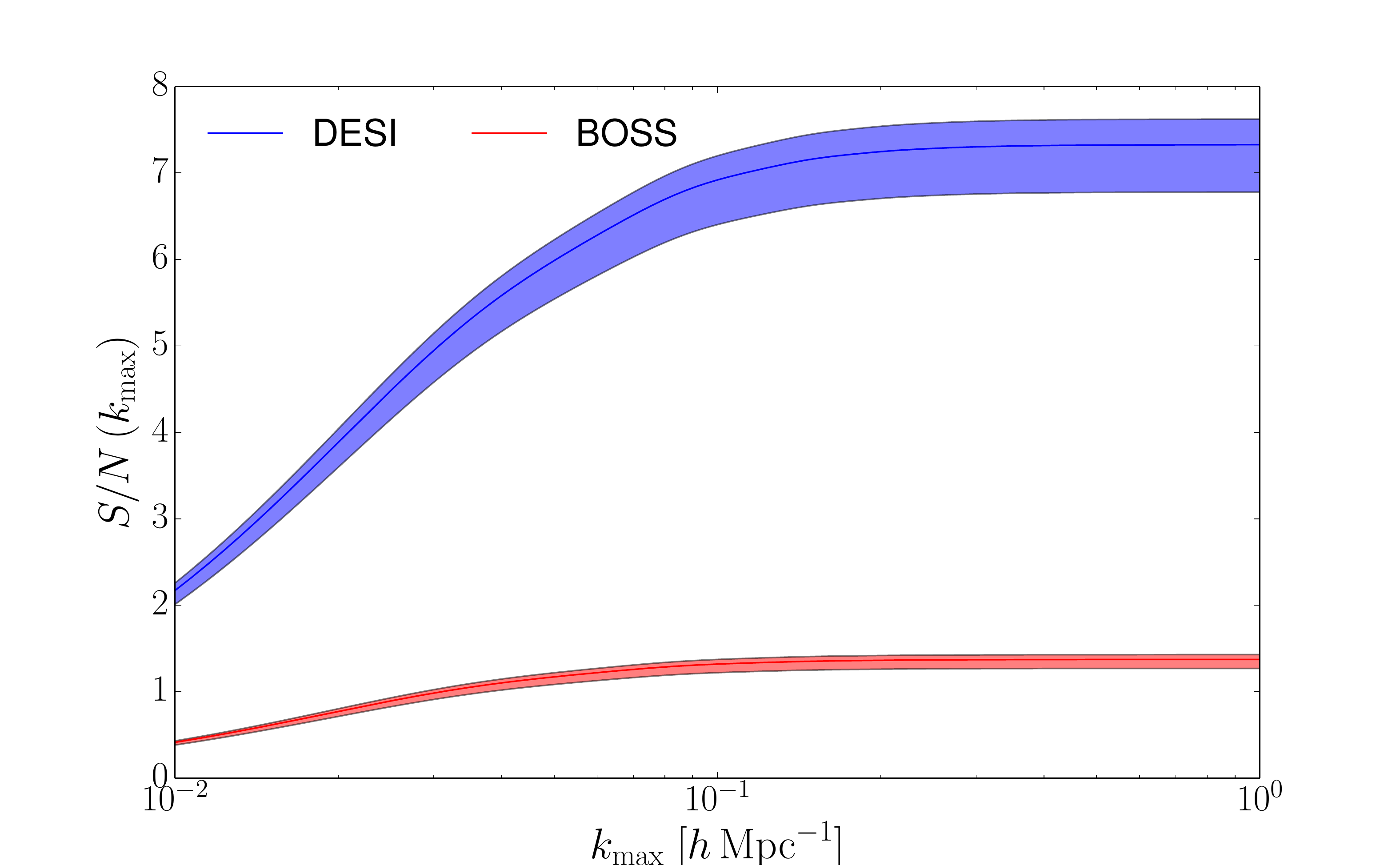}
  \caption{
    The plot shows the signal-to-noise ratio of the imaginary part of
    the QSO-\lya\ cross-power for two different
    surveys: BOSS (in red) and DESI (in blue). The shaded regions for
    each survey show variability in the $S/N$ ratio due to
    uncertainties in evolution/magnification biases for both QSOs and \lya (see text). The $S/N$ ratio is
    plotted as a function of the small-scale cutoff ($k_{\text{max}}$)
    to which the data-analysis of the survey extends. Given that the ratio grows with the
    number of small scale modes and then reaches a saturation point
    suggests that the $S/N$ of the relativistic effects is dominated
    by the shot-noise, and not by the cosmic variance. However, no
    systematic effects have been included in our analysis, and it is
    likely that in the real data, the signal is dominated by the
    systematics (see text for details). 
  }
  \label{fig:SN}
\end{figure}

We plot the signal-to-noise as a function of the small-scale cutoff
$k_{\text{max}}$ in Fig.~\ref{fig:SN}. We are showing the results for
current BOSS survey (in red) and future DESI survey (in blue). 

The
shaded regions represent the uncertainty in the $S/N$ ratio, when
allowing for the values of evolution and magnification biases to vary
($f_{\mathrm{evo}}^{\elya}$, $f_{\mathrm{evo}}^Q$ and $s$). The values were taken to be the same as in
Fig.~\ref{fig:asym_bias} and~\ref{fig:asym_fevolya}. The boundary
lines of the shaded region correspond to $+5\%$
($f_{\mathrm{evo}}^{\elya} = -3.15$, $f_{\mathrm{evo}}^Q = 6.19163$,
$s = 0.256883$ -
top line) and $-5\%$ ($f_{\mathrm{evo}}^{\elya} = -2.85$,
$f_{\mathrm{evo}}^Q = 4.74261$, $s = 0.450115$ -
bottom line) changes in evolution biases. The above values were taken
to be the same
for both BOSS and DESI surveys. While the uncertainty of the values for evolution
biases (or other bias factors) remains a source of a systematic error
for the future measurements, it does not significantly change the
ability to measure the relativistic effects.

The values for the noise power estimates used for the BOSS survey were
obtained using the relations presented in~\cite{mcquinn11,font14},
where the \lya\ noise power is given by the one-dimensional power
spectrum multiplied by the effective density of lines of sight per
unit area $n_{\text{eff}}$:
\be
N_\alpha^{\text{BOSS}} = P_\alpha^{1D}(k\mu) n_{\text{eff}}^{-1}.
\ee

The one-dimensional \lya\ power spectrum was obtained by integration
of $P_\alpha$ given by Eq.~(\ref{P_auto}). The value of density of lines
of sight was estimated for BOSS to be $n_{\text{eff}} = 10^{-3}\;
\left(h^{-1}\text{Mpc}\right)^{-2}$. Evaluating at a typical scale and
angle ($k = 0.14\;h\,\text{Mpc}^{-1}$,$\mu = k_\parallel/k = 0.6$) we
acquire the value of $N_\alpha^{\text{BOSS}} = 2670 \;\left(h^{-1}\text{Mpc}\right)^3$.
For the QSO noise we assume that it is given by the number density of
the QSOs in the survey. For BOSS we use the values of $164,017$ number of
quasars over $8,976$ square degrees, in the redshift range of $z = 2 -
3.5$.

For DESI we use the data presented in \cite{font14_DESI}. The noise
for the \lya\ forest is estimated to be 
$n P_\alpha(k=0.14\;h\,\mathrm{Mpc}^{-1},\mu=0.6) = 0.09$, which gives
$N_\alpha^{\text{DESI}} = 272 \;\left(h^{-1}\text{Mpc}\right)^3$. The
quasar noise was estimated using the table of QSO redshift
distribution, in the redshift range $z=1.96 - 3.55$, resulting in
$787,227$ quasars over $14,000$ square degrees.

The results shown in Fig.~\ref{fig:SN} show that in both surveys the
signal of the relativistic effects is shot-noise dominated, with both
achieving their maximum values at around 
$k_{\text{max}} = 0.1\;h\,\mathrm{Mpc}^{-1}$. The value for BOSS
survey is low, $S/N \sim 1.4$, indicating that the relativistic
effects cannot be measured with high significance, but they would
still be an important systematic effect. It is also likely that they
might be identified as continuum systematic effects, with which they
are likely degenerate to some extent (see the following section for
details). 

For DESI however, the results are much more promising, giving 
$S/N \sim 7.2$, suggesting that future surveys would be able to see
this effect. However, we caution the reader that this analysis was
very simple, especially in the sense that it did not involve any
systematic effects. Measurements of the BOSS
QSO-\lya\ cross-correlations (\cite{font13,font14}) have shown that
the uncertainties are dominated by the systematic effects, which need
to be carefully modeled. We claim however, that relativistic effects
need to be included in such analysis as they are an important
correction to the over signal.

\subsection{Discussion on systematic effects}
\label{sec:systematic}

Two of the major concerns about measuring the relativistic effects are
the possible degeneracy with other physical processes (e.g. primoridal
non-gaussianity, UV background fluctuations), and the effect of
systematic uncertainties in the data induced by the data-analysis
process (e.g. continuum fitting, redshift errors). In this section we
characterize the main sources of these uncertainties and discuss on
their possible impact on the results of our paper.

The largest source of systematic uncertainties comes from the
inability to clearly separate the QSO intrinsic continuum level and
mean absorption (and its fluctuations) in the \lya\ forest
region. The exact systematic effects from the continuum fitting depend
on the data-analysis technique as well as the continuum estimation
method, and would require a very careful survey-specific analysis to
carefully check what are the targeted uncertainties allowed in the
continuum fitting to be able to measure the relativistic effects presented in
this paper. This is well beyond the scope of this paper.

However, in the following we present a rough estimation of the
required continuum estimations. The continuum fitting procedure
affects the measured QSO-\lya\ cross-correlation because of the
inability to clearly separate the continuum and \lya\ forest large
scale fluctuations. This manifests itself as a scale dependent
continuum fluctuation dominated by large scales. These distortion
effects have been considered before (see \cite{font14}) and may
potentially mimic the relativistic signal. The theoretical models
used to account for such distortions (\cite{font14,font12DLA}) predict
an asymmetric distortion to the cross-correlation function with slight
dependence on the coordinate along the line of sight. However the
values are much below the relativistic signals.

On the other hand successive measurements of the BOSS
QSO-\lya\ cross-correlation (\cite{font13,font14}) have shown that the
final uncertainties are strongly dominated by the continuum systematic
effects. In a very simple picture where the continuum uncertainties
come from a mere wrong normalization of the continuum (i.e.~zero-mode
of the continuum fluctuations) it would produce a multiplicative
offset in the cross-correlation function. As the relativistic
corrections seem to apply a nearly constant distortion above $\sim
40\;h^{-1}\mathrm{Mpc}$, they can be mimicked by a wrong continuum
level. Whereas the relativistic effects are asymmetric, a mere wrong
continuum level would introduce a symmetric offset. As a rough
estimation the continuum level in the future surveys must be known to
a better than a few percent level, if a high-significance measurement
of the relativistic effects is to be achieved. 

  As shown in \cite{font13}, the quasar redshift errors
  would shift the whole correlation function for a constant amount, 
  while also smearing the small scales due to the
  variance of the redshift estimates. A systematic shift of the
  redshift of the quasars would imply an incorrect wavelength
  callibration of the quasar spectra. This would in turn mean that the
  redshifts of the \lya\ forest are also systematically offset for the
  same amount ($\Delta_z$). Combining the above generates a
  coordinate translation of the kind $\langle
  \Delta_Q(z_1)\delta_F(z_2)\rangle \rightarrow \langle
  \Delta_Q(z_1- \Delta_z)\delta_F(z_2 - \Delta_z)\rangle$. This is
  exactly how the effect was modelled in \cite{font13},
  where the theoretical correlation function was convolved with a
  gaussian profile, with the mean corresponding to $\Delta_z$ and a
  width corresponding to the variance of redshift estimates. Such a
  translation would only shift the whole cross-correlation
  function, so that its peak would be at $\Delta_z$ and not at
  $z_2-z_1=0$. The shape would not be changed (except on small scales
  due to the smearing of the kernel function), and would not change
  the signal of the asymmetric cross-correlation function.

Another important systematic effect inherent to data-analysis of the
\lya\ forest is the inability to properly filter out metal contaminants
that are falsely identified as \lya\ absorbers. While most of these
metal contaminants only affect small scales of the \lya\ forest
fluctuations, there is one metal line very close to the \lya\ line,
that is known to be associated with Si III absorption. First observed
and modeled in \cite{mcdonald06}, it was shown that it causes a
smaller, secondary maximum in the \lya\ auto-correlation
function. However, in the QSO-\lya\ cross-correlation function such an
effect would undoubtedly induce asymmetry in the cross-correlation
function. To assess this systematic signal we adopt a simple model used
in \cite{mcdonald06}, where the contribution of the Si III to the
total flux fluctuations can be modeled as shifted and rescaled
\lya\ fluctuations, $\delta_F(z) = \delta_\alpha(z) + a
\delta_{\alpha}(z + z_3)$, where the shift due to Si II is $z_3 =
v_3 \sim 0.0078$, and the scale was estimated to be $a \sim 0.044$
at $z=2.5$. To the first order in $z_3$, the correction to the overall
cross-correlation function is equal to $\langle \Delta_Q \delta_F
\rangle = \langle \Delta_Q \delta_\alpha \rangle + a \langle \Delta_Q
\delta_\alpha \rangle + a z_3 \langle \Delta_Q \partial_z\delta_\alpha
\rangle$. We see that the leading (zero-th order) correction will
scale with $a$, and thus be suppressed by more than an order of
magnitude compared to the \lya\ signal and is thus a negligible
effect to our results.

One of the physical processes that might modify the behaviour of large
scale fluctuations is the elusive primordial non-gaussianity, that has
been extensively searched for in both galaxy surveys and cosmic
microwave data \cite{boss13nongauss,planck15nongauss}. The models in the literature so far
have focused mainly on leading order effect on the density
fluctuations, where they have shown that $f_{NL}\neq 0$ generates a
scale dependent density bias factor, that has a correction of the
order of $(\HH/k)^2$ to the observed over-density. This is of the
same order of magnitude as our potential terms, and moreover, it is
symmetric, and will thus no affect the asymmetric correlation
function. Similar behaviour can be observed for the effect of UV
background fluctuations. The current models (\cite{gontcho14,pontzen14}) show that such
fluctuations change only the symmetric part (adding power only to the real
part of the power spectrum). However, some second
order corrections are expected from cross-correlating these symmetric
effects with the leading asymmetric terms, and a more detailed
analysis would be required to study the combined effects. 

\section{Conclusions}
\label{sec:conc}

In this paper we have computed the relativistic effects in the
\lya\ forest in the linear regime on large scales. There are two main
reasons to conduct such a study: firstly, the future surveys are going
to measure the regime of large scales where the relativistic effects may
become important ($\sim 10\%$ of the signal at the BAO scale), 
and secondly, the already existing data can be used
with multiple tracers to boost specific signals beyond the standard
Newtonian terms and constrain cosmological information.

Using two differently biased tracers to boost the small signal of the
relativistic effects have been suggested in the
past~\cite{McDonald:2009ud,Yoo:2012se,Bonvin:2013ogt,Alonso:2015sfa,Fonseca:2015laa}
but mostly in the context of different galaxy populations. Here we
have adopted the approach, used the newly
computed
relativistic expansion for the \lya\ forest and shown that when
combining two tracers with very different bias factors (such as
Quasars and \lya\ forest) one can enhance the desired signal. 

The main effect on the cross-correlation function can be measured
through an anti-symmetric part of the correlation function. We show that
the relativistic effects, and most prominently the Doppler term, give
rise to the most dominant contribution. While these new terms include
a variety of exciting information, there are two sub-dominant effects
that need to be carefully modeled: the redshift evolution of the
density (and velocity-gradient) bias factors of the tracers, and the
QSO lensing term. In the light of the lensing effects, the
cosmological signal is strongly contaminated by the magnification, and
to lesser extent, evolution bias of the QSOs. While all the bias
parameters can be determined from the independent surveys it is
important to note that they can mimic the effects of different
cosmology or theory of gravity.

For the given configuration of the tracers the results show that the
contaminating effects are surprisingly small at BAO scale. While this
only reflects the specific values of the bias parameters and their
redshift evolution for these two tracers, it can be used to determine
the effect of the Doppler terms at the BAO scale.

Another possible advantage for the precision measurements of future
surveys is to isolate the anti-symmetric signal of lensing or
potential terms and
use them to test the validity of the Einstein's General Relativity or
constrain the modifications of the theory of gravity. The dominant
signal, coming from the Doppler term, would not change under a
different theory of gravity, unless the theory violates the
energy-momentum tensor conservation. Even though, the potential and
lensing signals are small, if
measured they would provide a unique test of General Relativity on the
BAO scales, thus in the regime where perturbations can be measured
much better than at horizon scales.

\section*{Acknowledgement}
We thank An\v{z}e Slosar and Ruth Durrer for fruitful
discussions. We also thank anonymous referee for insightful
suggestions.
ED and MV are supported by the ERC grant 'cosmoIGM' and by INFN/PD51 INDARK grant.


\appendix
\section{Angular power spectrum}
\label{sec:cl}

After having computed the 2-point correlation function we can compute the power spectrum as well. Since the observables defined in Eqs.~(\ref{Lya_Delta},\ref{delta_Q}) depend on angular positions and redshifts, spherical harmonics are the natural basis to expand them as follows
\bea
\Delta_A \left( z, \bn \right) = \sum_{\ell m} a^A_{\ell m } \left( z \right) Y_{\ell m} \left( \bn \right) \, , \\
a^A_{\ell m } \left( z \right) = \int d\Omega_\bn \Delta_A \left( z, \bn \right)  Y^*_{\ell m } \left( \bn \right) \, .
\eea
Then, we define the redshift dependent angular power spectra
\be
C^{AB}_\ell \left( z , z' \right) = \langle a^A_{\ell m } \left( z \right) a^{B^*}_{\ell m } \left( z' \right) \rangle
\ee
where $A$ and $B$ denotes two observables. It is convenient to write the angular power spectra in terms of integral over the angular transfer function $\Delta_\ell^A \left( \eta, k \right)$ through
\be
C^{AB}_\ell \left( z , z' \right) = 4 \pi \int \frac{dk}{k} \Delta_\ell^A \left(  k,\eta \right) \Delta_\ell^{B^*} \left(  k, \eta' \right) \mathcal{P}_R \left( k \right)\, ,
\ee
where $\mathcal{P}_R \left( k \right)= k^3 P_R \left( k \right) /(2\pi^2)$ is the dimensionless primordial curvature power spectrum. The same formalism has been successfully applied to Cosmic Microwave Background (CMB), Weak Lensing (WL) and galaxy clustering. Because of that, the implementation in a Boltzmann code is straightforward. In this work we have modified the public \class{} code~\cite{Lesgourgues:2011re,Blas:2011rf}, in particular the part included in \classgal{}~\cite{DiDio:2013bqa} and we have adopted it to computed, through Eq.~(\ref{2point_cl}), the correlation function of non-local terms.

As shown in galaxy clustering, the analysis based on redshift dependent 2-dimensional power spectra recovers the same amount of information of the traditional 3-dimensional analysis, see e.g.~\cite{Asorey:2012rd,DiDio:2013sea,Nicola:2014bma}. In addition, it is the natural way to include relativistic or light-cone effects and it has been already considered in several works. Hence, the formalism here discussed can be used directly to perform a 3-dimensional analysis, including all the relativistic effects to first order in perturbation theory.

\subsection*{Ly-$\alpha$ power transfer function}
For the transmitted flux, from Eqs.~(\ref{DeltaF_linear}, \ref{tau_fin}), we compute the angular transfer function
\bea
\Delta_\ell \left(  k, \eta \right) &=& b T_D j_\ell \left( k r \right)  + b_v \HH^{-1} T_\Theta j''_\ell \left( k r \right) 
\nonumber \\
&&
+ b_R
  \left\{
   \left[  \left( 3  + \frac{\dot \HH}{\HH^2 } - f_{\rm evo}^{\elya}\right) T_\Psi + \HH^{-1} T_{\dot \Phi} + \left(  f_{\rm evo}^{\elya}-3  \right)  \frac{\HH}{k^2} T_\Theta \right] j_\ell \left( k r \right) 
 \right.
\nonumber \\
&& 
\left.
+
\left(  3 + \frac{\dot \HH}{\HH^2 }  - f_{\rm evo}^{\elya} \right) k^{-1} T_\Theta j'_\ell \left( k r \right)  +
\left( 2  + \frac{\dot \HH}{\HH^2 }  - f_{\rm evo}^{\elya} \right) \int_0^r T_{\dot \Psi + \dot \Phi} j_\ell \left( k r' \right) dr'  \right\} \, ,
\nonumber \\
\eea
where a prime denotes the derivative of the spherical Bessel functions with respect to their argument. All the quantities are evaluated at $\left(k, \eta \right)$ if not written explicitly.
\subsection*{Quasars transfer function}
Following the notation of~\cite{DiDio:2013bqa}, we reproduce here the angular transfer function for QSO
\bea
&&\hspace*{-4mm}\De_\ell(z,k) = 
 j_\ell(kr ) \left[ b T_D + \left(\frac{\dot{\cal H}}{{\cal H}^2}+\frac{2-5s}{r {\cal H}} +5s - f^Q_{\rm evo} +1 \right) T_\Psi \nonumber \right. \\
&&\left. \hspace{1.5cm}+~\left(  -2 +5s \right) T_\Phi + {\cal H}^{-1} T_{\dot{\Phi}} \right]  \nonumber \\
&&+ \left[ \frac{d j_\ell}{dx}(kr) \!\left( \frac{\dot{\cal H}}{{\cal H}^2} + \frac{2 -5s}{r {\cal H}} +5s\!- \!f^Q_{\rm evo} \right)+ \frac{d^2 j_\ell}{dx^2}(kr) \frac{k}{\cal H} \left( { f^Q_\text{evo}}-3 \right) j_\ell(kr)\frac{\HH}{k}\right]  T_V
 \nonumber \\
&&+ \int_0^{r(z)} \hspace*{-0.4cm} dr \, j_\ell(kr) \left[  T_{{\Phi}+ \Psi} \left( \frac{2-5s}{2}\right)\left(\!\ell(\ell+1) \frac{r(z)-r}{r(z) r} + \frac{2}{r(z)} \right) \right. \nonumber \\
&&\hspace{25mm}
+ \left.  T_{\dot\Phi+\dot\Psi}  \left(\!\frac{\dot{\cal H}}{{\cal H}^2} + \frac{2-5s}{r(z) {\cal H}} +5s -f^Q_{\rm evo}\!\right)_{r(z)}
\right]\, .
\label{Fl}
 \eea

\section{CLASS modifications}
\label{sec:CLASS}
Here we present the modified transfer function in \class{}
code~\cite{Lesgourgues:2011re,Blas:2011rf}, in particular for the
galaxy number counts calculation developed in
\classgal{}~\cite{DiDio:2013bqa}, where the angular power spectrum between the $i$-th and the $j$-th redshift bins is denoted by
\be
c^{ij}_\ell = 4 \pi \int \frac{dk}{k} \mathcal{P_R} \left( k \right) \Delta_\ell^i \left( k \right) \Delta_\ell^j \left( k \right) 
\ee
with 
\bea
\Delta_\ell \left( k \right)&=&\Delta_{\ell}^{\mathrm{Den}_i}\hspace{4.5cm} \text{(Density)}
\nonumber \\
&&
+\Delta_{\ell}^{\mathrm{Red}_i}	\hspace{4.2cm}\text{(Redshift space distortion)}
\nonumber \\
&&
+ \Delta_{\ell}^{\mathrm{D}1_i} +\Delta_{\ell}^{\mathrm{D}2_i}\hspace{3cm}\text{(Doppler)}
\nonumber \\
&&
+ \Delta_{\ell}^{\mathrm{G}1_i}+\Delta_{\ell}^{\mathrm{G}2_i}+\Delta_{\ell}^{\mathrm{G}3_i}+\Delta_{\ell}^{\mathrm{G}5_i}\hspace{0.4cm}\text{(Gravitational potential)}
\eea
and
\begin{eqnarray}
\Delta_{\ell}^{\mathrm{Den}_i} &=& \int_0^{\eta_0} d\eta W_i \, b S_\mathrm{D}  \, j_\ell \nonumber \\
\Delta_{\ell}^{\mathrm{Red}_i} &=& \int_0^{\eta_0} d\eta \, W_i \frac{b_v}{aH}  S_{\Theta} \, \frac{d^2j_\ell}{dx^2} \nonumber \\
\Delta_{\ell}^{\mathrm{D}1_i} &=&  \int_0^{\eta_0} d\eta \, W_i  \frac{ 4 +  \frac{\dot H}{aH^2}  - f_{\rm evo}^{\elya} }{k} S_{\Theta} b_R \, \frac{d j_\ell}{dx}  \nonumber \\
\Delta_{\ell}^{\mathrm{D}2_i} &=& \int_0^{\eta_0} d\eta \, W_i   \left( f_{\rm evo}^{\elya} - 3\right)  \frac{aH}{k^2} S_{\Theta}b_R  \, j_\ell \nonumber \\
 \Delta_{\ell}^{\mathrm{G}1_i} &=& \int_0^{\eta_0} d\eta \, W_i \,  S_\Psi b_R \, j_\ell \nonumber \\
 \Delta_{\ell}^{\mathrm{G}2_i} &=&  - \int_0^{\eta_0} d\eta \, W_i \, \left( 3 + \frac{\dot H}{aH^2} - f_{\rm evo}^{\elya}\right) S_{\Phi} b_R \, j_\ell \nonumber \\
\Delta_{\ell}^{\mathrm{G}3_i} &=& \int_0^{\eta_0} d\eta \, W_i \, \frac{1}{aH} S_{\dot\Phi}b_R  \, j_\ell \nonumber \\
 \Delta_{\ell}^{\mathrm{G}5_i} &=& \int_0^{\eta_0} d\eta \, W_i^{\mathrm{G}5} \, S_{(\Phi+\Psi)} k  \, \frac{d j_\ell}{dx} ~,
\label{eq:delta_terms}
\end{eqnarray}
where $W_i$ denotes the selection function of the $i$-th redshift bin of the survey.
We have omitted all the arguments: $k$ for the transfer functions, $(\eta,k)$ for the source functions, $x\equiv k(\eta_0-\eta)$ for the Bessel functions, and $\eta$ for selection and background functions. For the integrated term G5, we have defined
\begin{eqnarray}
W_i^{\mathrm{G}5}(\eta) &=& \int_0^\eta \!\! d\tilde{\eta}b_R  W_i(\tilde{\eta}) \left( { 3} +\frac{\dot H}{aH^2} - f_{\rm evo}^{\elya}\right)_{\tilde{\eta}}~. \nonumber
\end{eqnarray}
The term G4 vanishes because of the absence of time-delay effect in \lya\ observable.

\bibliographystyle{JHEP}
\bibliography{biblio_Lya}
\end{document}